\documentclass[20pt,preprint]{aastex}

\usepackage{natbib}
\usepackage{multicol}
\usepackage{color}
\usepackage{multirow}

\begin{document}

\title{Plasma diagnostic potential of 2p4f in N$^+$ -- accurate wavelengths and oscillator strengths}


\author{\rm  X{\footnotesize IAOZHI}  S{\footnotesize HEN}$^{1,2}$ , J{\footnotesize IGUANG} L{\footnotesize I}$^{3\dag}$, P{\footnotesize ER} J{\footnotesize \"{O}NSSON}$^{4}$, J{\footnotesize IANGUO} W{\footnotesize ANG}$^{3}$ }
\affil{\footnotesize
$^1$ School of Physics Science and Nuclear Energy Engineering, Beihang University, Beijing 100191, China;\\
$^2$ School of Mechanical and Electrical engineering, Handan College, Handan 056005, China;\\
$^3$ Data Center for High Energy Density Physics, Institute of Applied Physics and Computational Mathematics, PO Box 8009, Beijing 100088, China;\\
       Corresponding author: Li\_Jiguang@iapcm.ac.cn\\
$^4$ Materials Science and Applied Mathematics, Malm\"{o} University, 205 06 Malm\"{o}, Sweden
}

\begin{abstract}
Radiative emission lines from nitrogen and its ions are often observed in nebulae spectra, where the N$^{2+}$ abundance can be inferred from lines of the 2p4f configuration.
In addition, intensity ratios between lines of the 2p3p -- 2p3s and 2p4f -- 2p3d transition arrays can serve as temperature diagnostics.
To aid abundance determinations and plasma diagnostics, wavelengths and oscillator strengths were calculated with high-precision for electric-dipole (E1) transitions from levels in the 2p4f configuration of N$^{+}$.
Electron correlation and relativistic effects, including the Breit interaction, were systematically taken into account  within the framework of the multiconfiguration Dirac-Hartree-Fock (MCDHF) method.
Except for the 2p4f - 2p4d transitions with quite large wavelengths and the two-electron-one-photon 2p4f -2s2p$^3$ transitions, the uncertainties of the present calculations were controlled to within 3\% and 5\% for wavelengths and oscillator strengths, respectively.
We also compared our results with other theoretical and experimental values when available.
Discrepancies were found between our calculations and previous calculations due to the neglect of relativistic effects in the latter.
%

\end{abstract}



\section{INTRODUCTION}

Nitrogen is one of the most abundant elements in the universe.
Radiative emission lines from nitrogen and its ions are often observed in nebulae spectra,
and some of the lines are suitable for abundance determinations and plasma diagnostics~\citep{Liu00,Fang11}.
In particular, there has been a great interest in lines originating from levels in the 2p4f configuration of N$^+$.
For example, Liu et al. determined the N$^{2+}$/H$^+$ ion abundance in NGC 6153 using the line intensities of the 2p4f -- 2p3d transitions~\citep{Liu00}.
%
%
A similar determination was done in the Orion nebula by Escalante and Morisset who pointed out that a major concern is the uncertainty in the line fractions involving the 2p4f
term, where LS-coupling is not a good approximation~\citep{Escalante05}. 
%
%
Fang et al. demonstrated that the intensity ratios between the 2p3p $^3$D -- 2p3s $^3$P$^o$ and 2p4f G(9/2) -- 2p3d $^3$F$^o$ transitions have a relatively strong temperature dependence, and thus can serve as a temperature diagnostics~\citep{Fang11}.
In addition, there exist a few lines from the 2p4f configuration in lightning~\citep{Wallace63}, which play key roles in the determination of properties such as temperature and pressure~\citep{Prueitt63, Uman64}.
%

Accurate atomic parameters for the transitions from the 2p4f configuration are still scarce, although they are important for abundance determinations and plasma diagnostics as mentioned earlier.
Mar et al. reported experimental probabilities for 20 transitions between the 2p4f and 2p3d configurations of the N$^{+}$ ion produced in a pulsed discharge lamp containing helium and nitrogen gas.
However, the absolute rates were obtained by using data available in the literature as a reference~\citep{Mar00}.
In addition, some experiments were carried out for measuring lifetimes of levels belonging to the 2p4f configuration~\citep{Denis68, Pinnington70, Brink78, Desesquelles71, Fink68, Warren71}.
Yet, it is sometimes difficult to infer transition rates through lifetimes since there are always several decay channels from an individual level.
Turning to theory, Kelly reported values of the single-electron integrals for the 2p4f -- 2p3d transitions in the Hartree-Fock-Slater approximation~\citep{Kelly64}.
Based on these data, Wiese et al. later calculated the corresponding oscillator strengths~\citep{Wiese65}.
Victor and Escalante also obtained atomic parameters for the 2p4f -- 2p3d and 2p4f -- 2p4d transitions using a model potential method~\citep{Victor88}.
%
Finally, as part of the Opacity Project, oscillator strengths involving the 2p4f configuration were calculated using the R-matrix method~\citep{OP95}.  
However, relativistic effects were neglected in this calculation, resulting in relatively large uncertainties for the atomic parameters.

\begin{figure}[!th]
\centering
\includegraphics[width=13cm]{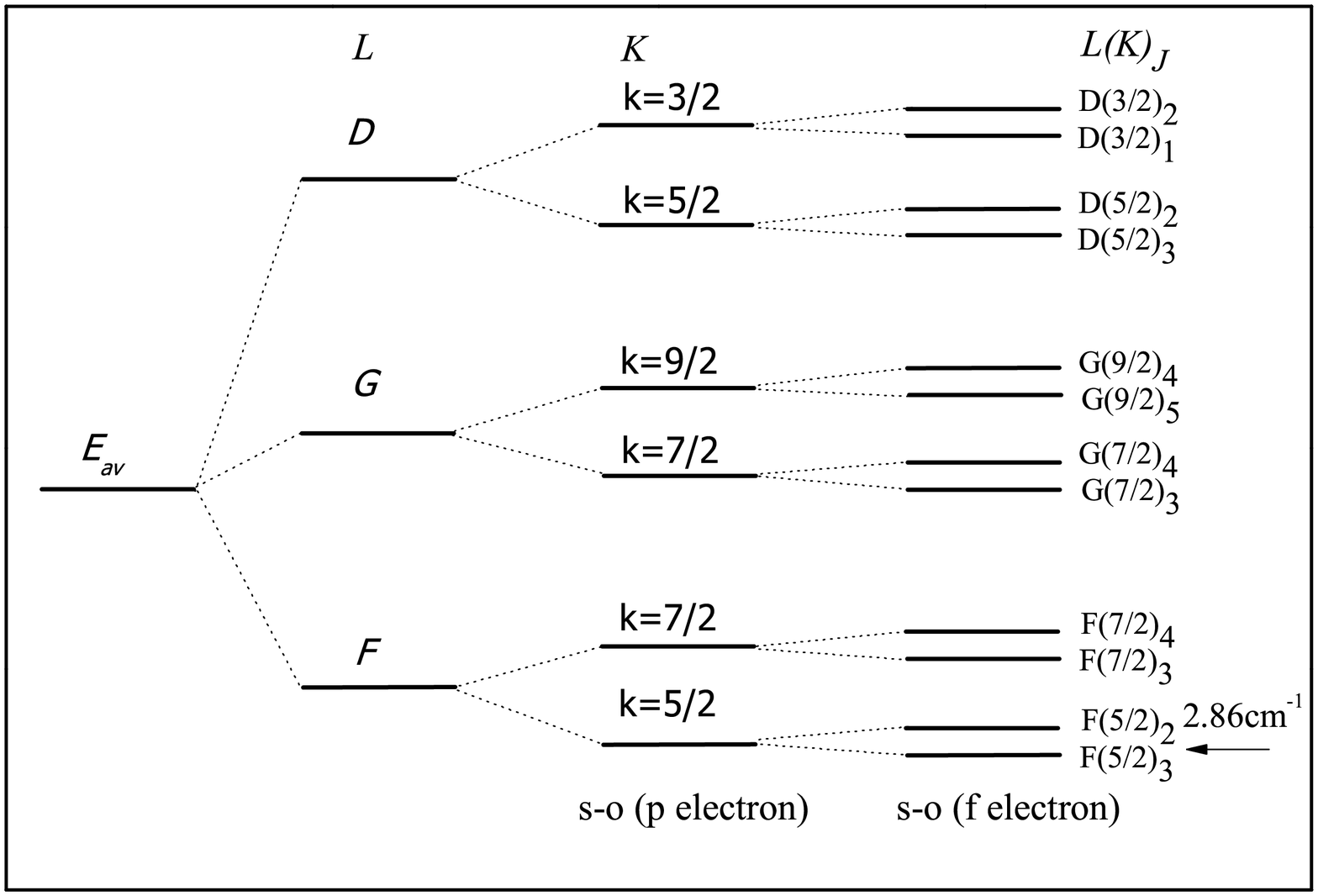}
\caption{\label{fig:Fig1}
The energy level structure of the 2p4f configuration.
E$_{av}$ is the configuration average energy.
The largest interaction -- the spin-independent portion of the electron-electron Coulomb interaction gives rise to three terms F, G and D.
The spin-orbit (s-o) interaction of the 2p electron is the second most important interaction, and produces a separation according to the two possible values $K=L \pm s_p (s_p=1/2)$.
The s-o interaction of 4f electron brings about very small splittings. }
\end{figure}

Because of the weak spin-dependent Coulomb interaction between the 2p and 4f electrons and the small spin-orbital interaction for the 4f electron itself, the level structure in the 2p4f configuration is best described in LK-coupling~\citep{Cowan81}.
Also, fine-structure splittings in this configuration are extremely small.
For example, the separation between the F(5/2)$_3$ and F(5/2)$_2$ levels is just 2.86 cm$^{-1}$ as shown in Fig.~\ref{fig:Fig1}.
To describe this level structure, it is essential to accurately capture both relativistic and electron correlation effects.
Improving on our previous work on transition probabilities from the 2p4f configuration~\citep{Shen10}, in which a simple correlation model was adopted, we performed large-scale calculations using the multiconfiguration Dirac-Hartree-Fock (MCDHF) method.
A multireference active set approach was utilized to systematically generate the configuration space~\citep{Sturesson07}.
In particular, higher-order electron correlation effects were taken into account by means of an extended set of configurations in the multireference ~\citep{Li12}.
In addition, we also considered the Breit interaction -- the main relativistic correction to electron interactions~\citep{Grant07}.
The uncertainties of the present calculations were controlled to within 3\% for wavelengths and to about 5\% for oscillator strengths of most of lines, respectively.
Based on the present work, we evaluated previous theoretical results and found some discrepancies owing to the neglect of relativistic effects in previous calculations.

\section{THEORETICAL METHOD AND COMPUTATIONAL
MODEL}

\subsection{\label{app:subsec} Theoretical method}                                  

We employed the multiconfiguration Dirac-Hartree-Fock (MCDHF) method to calculate the atomic state wave functions (ASFs).
The details of the method are described in the monograph by Grant~\citep{Grant07} and here we just give a brief account.
%
%

In the MCDHF method the ASFs are linear combinations of symmetry adapted configuration state functions
(CSFs) with the same parity $P$, angular momentum $J$, and its $M_J$ component along $z$ direction
\begin{equation}
\Psi(P J M_J )=\sum^{N_{CSFs}}_{k=1} c_{k} \Phi(\gamma_k P J M_J ).
\end{equation}
In the expression above $c_{k}$ are the expansion  coefficients and $\gamma_k$ denote other
appropriate labeling of the CSFs, e.g. orbital occupation numbers and coupling trees.
The CSFs are built from products of one-electron Dirac orbitals.
In the self-consistent field (SCF) procedure, both the radial parts of the Dirac orbitals and the
expansion coefficients are determined to minimize the energies based on the Dirac-Coulomb Hamiltonian.
Calculations can be performed for a single level, but also for a portion of a spectrum in an extended optimal level
(EOL) scheme, where the minimization is on a weighted sum of energies.
The Breit interaction
between all electron pairs is included in subsequent relativistic configuration interaction (RCI) calculations, where the radial orbitals are fixed and only the expansion coefficients are optimized \citep{Grant80}.

For a transition between an initial $i$ and a final $f$ state the transition parameters such as the weighted oscillator strength $gf$ and the transition rate $A$ 
can be expressed in terms of the reduced matrix element
\begin{eqnarray}
\langle \Psi_i \| O^{(L)} \| \Psi_f \rangle^2,
\end{eqnarray}
where $O^{(L)}$  is the multipole radiation field operator.
A biorthogonal transformation technique is adopted to relax the restrictions from standard Racah algebra
so that the initial and final state ASFs can be built from the different radial orbital sets~\citep{Olsen95}.
All calculations were performed using the GRASP2K package~\citep{Per13} which is the latest version of GRASP~\citep{Grant80}.
%

\subsection{\label{app:subsec} Computational model}                                  

The accuracy of MCDHF and RCI calculations is to a large extent determined by the CSF expansions. 
In this work, the active set approach was adopted to generate the CSF expansions.
%
%
Calculations were done by parity, meaning that states of the even and odd parity, respectively, were optimized separately.
Based on the experience from our previous work \citep{Shen10} the reference configurations
2s$^2$2p$^2$; 2s$^2$2p3p; 2s$^2$2p4p; 2s2p$^2$3s; 2s$^2$2p4f and 2s2p$^3$; 2s$^2$2p3s; 2s$^2$2p3d;
2s$^2$2p4s; 2s$^2$2p4d;2s$^2$2p5s were chosen for the two parities.
It is worth noting that the higher-order electron correlations can be accounted for through an extended set of reference configurations.
The CSFs were formed from all configurations that could be obtained by replacing the occupied orbitals in the reference configurations 
with orbitals in an active set according to some rules. The rule together with the active space define the computational model. In this work 
we allowed single (S) and double (D) replacements from the valence orbitals as well as from the valence and the 1s core orbitals and the models were denoted
nSDV and nSDC, where n indicate the maximum principal quantum number of the orbitals in the active set.   
The orbitals in the active set were augmented layer by layer so as to be able to monitor the convergence of the physical quantities concerned.
The number of CSFs is displayed in Table~\ref{tab:table1} as a function of the computational model.

%
Due to convergence problems in the self-consistent calculation for the even parity reference configurations, we added the following configurations
2s$^2$3d$^2$, 2s2p$^2$3d, 2s2p3p3d, 2s3s3d$^2$, 2p$^4$, 2p$^3$3p, 2p$^2$3s3d to stabilize the calculation.
This first step was labeled with DF in Table~\ref{tab:table1} only for convenience.
As the active set of orbitals was enlarged, only the orbitals in the added layer were optimized.
The final calculations allowing for substitutions also from the 1s core orbital were done in RCI. For these calculations the Breit interaction was included as well.
%
%
%
%


\begin{deluxetable}{lllrrrrrr}
\tabletypesize{\scriptsize}
\tablecaption{\label{tab:table1}
{\footnotesize
The number of CSFs ($N_{CSFs}$) with different symmetries of the angular momentum (J) and the parity in different computational models.
AS denotes the highest principal quantum number $n$ in the active set of orbitals.
DF stands for the calculations based on the CSFs of the reference configurations.
nSDV and nSDC denote the computational models.
%
%
}}
\tablewidth{0pt}
\tabcolsep 2pt
\tablehead{
&&&& \multicolumn{5}{c}{ \textit{N$_{CSFs}$}  }     \\
\cline{4-9}
\colhead{Reference Configuration}& \colhead{AS~~~}& \colhead{Model} &
\colhead{J=0}& \colhead{J=1}& \colhead{J=2}& \colhead{J=3}&
\colhead{J=4}& \colhead{J=5} }
\startdata
\multicolumn{9}{c}{Even}\\
$\{$2s$^2$2p$^2$; 2s$^2$2p3p; 2s$^2$2p4p; 2s2p$^2$3s; 2s$^2$2p4f; \\
2s$^2$3d$^2$; 2s2p$^2$3d; 2s2p3p3d; 2s3s3d$^2$; 2p$^4$; 2p$^3$3p; 2p$^2$3s3d$\}$ &                       & DF          & \textbf{41}       & \textbf{89}      & \textbf{106}      & \textbf{77}        & \textbf{42}      & \textbf{13} \\[0.2cm]
\{2s$^2$2p$^2$; 2s$^2$2p3p; 2s$^2$2p4p; 2s2p$^2$3s; 2s$^2$2p4f\}      & 4~~~~~             & 4SDV         &906     &2297    &3020    &2841     &2193   &1371  \\
                                                            &             5           & 5SDV         &3064    &8167    &11296   &11736    &10251  &7625  \\
                                                            &             6           & 6SDV         &7172    &19603   &28028   &30878    &29098  &23950  \\
                                                            &             7           & 7SDV         &13808   &38369   &56239   &64626    &64425  &57154  \\
                                                            &             7           & 7SDC         &71635   &200660  &294281  &339943   &339811 &303282  \\
                                                            &                         &             &        &        &        &         &       &       \\ \\
\multicolumn{9}{c}{Odd}\\
\{2s2p$^3$;2s$^2$2p3s;2s$^2$2p3d;2s$^2$2p4s;2s$^2$2p4d;2s$^2$2p5s\} &               &DF&6         &16      &15      &7        &2      &  \\
                                                            &  4~~~~~                    & 4SDV         &1033    &2727    &3463 	&3230      &2406   &  \\
                                                            &  5                    & 5SDV         &3035    &8255    &11231  &11606     &9917   &  \\
                                                            &  6                    & 6SDV         &7109    &19682   &27856  &30598     &28473  &  \\
                                                            &  7                    & 7SDV         &13609   &38147   &55462  &63516     &62654  &  \\
                                                            &  7                    & 7SDC         &68459   &192172  &280405 &322427    &319636 &   \\
                                                            &                       &            &        &        &        &        &       &  \\
\enddata
\end{deluxetable}

\section{RESULTS AND DISCUSSION}

\subsection{\label{app:subsec} Excitation energies and fine structure splittings }    

Excitation energies of levels in the 2p4f configuration, obtained with different computational models, are listed in the upper part of Table~\ref{tab:table2}.
The L[K]$_{\rm J}$ notation is used to mark these levels. For convenience we also present the LS notation.
It can be found from this table that correlation effects, not only between valence electrons, but also between the core and valence ones, are very important.
For example, excitation energies are reduced by about 6.5\% under 4SDV model, and further adjusted by about 400 cm$^{-1}$ when considering core-core and core-valence correlations in the 7SDC model.
The influence of the Breit interaction on the excitation energies is so small as to be negligible.
Comparing with experimental values from NIST we see that the uncertainties are less than 0.14\% for excitation energies of the 2p4f configuration.

As mentioned earlier, the level structure of the 2p4f configuration is best described in the LK-coupling scheme and the fine-structure splittings are only a few wave numbers.
Therefore, the calculated fine-structure splittings are indispensable physical quantities for judging the quality of the ASFs. In the lower part of Table~\ref{tab:table2}, we present the calculated splittings.
One should keep in mind that these calculations were performed within the fully relativistic framework.
In other words, the relativistic effects were considered from the start.
As a results, the discrepancies in fine-structure splittings at the DF level is attributed to the neglected electron correlation effects.
For instance, the order of the energy levels belonging to the F(5/2) term is not correct until the 5SDV model has been reached.
%
%
After including the Breit interaction, the calculated fine-structure splittings are in good agreement with the NIST values.
%


%
Excitation energies for levels in the 2s2p$^3$ and 2p3d configurations are reported in Table~\ref{tab:table3} as functions of the computational models.
A good agreement with the NIST values is found. The difference is overall smaller than $0.2\%$, except for the 2s2p$^3~^5$S$^o_2$, ~$^1$D$^o_2$, ~$^3$S$^o_1$ and $^1$P$^o_1$ states where the uncertainties approach $1\%$.
%

\begin{deluxetable}{lccccccccccccccccc}
\tabletypesize{\scriptsize}
\tablecaption{\label{tab:table2}
{\footnotesize
Excitation energies (in cm$^{-1}$) and fine-structure splittings (in cm$^{-1}$) of the 2p4f configuration from different computational models. 
%
%
%
%
}}
\tablewidth{0pt}
\tabcolsep 1.9pt
\tablehead{
\colhead{Model} & \colhead{  F(5/2)$_3$} & \colhead{  F(5/2)$_2$}
&  & \colhead{ F(7/2)$_3$} & \colhead{ F(7/2)$_4$}
&  & \colhead{ G(7/2)$_3$} & \colhead{ G(7/2)$_4$}
&  & \colhead{ G(9/2)$_5$} & \colhead{ G(9/2)$_4$}
&  & \colhead{ D(5/2)$_3$} & \colhead{ D(5/2)$_2$}
&  & \colhead{ D(3/2)$_1$} & \colhead{ D(3/2)$_2$}  \\
\cline{2-3} \cline{5-6} \cline{8-9} \cline{11-12} \cline{14-15} \cline{17-18}
 &($^1$F$_3$)&($^3$F$_2$)&&($^3$F$_3$)&($^3$F$_4$)&&($^3$G$_3$)& ($^3$G$_4$)&&($^3$G$_5$)&($^1$G$_4$)&&($^3$D$_3$)&($^3$D$_2$)&&($^3$D$_1$)& $^1$D$_2$)}
\startdata
\multicolumn{18}{c}{Excitation energies}\\
DF     &219061   &214782  &&219090  &219093  &&219337  &219345  &&219458  &219471  &&219473  &219064  &&214198 &219482   \\
4SDV    &205730   &205730  &&205756  &205763  &&205995  &206004  &&206112  &206121  &&206119  &206122  &&206202 &206210   \\
5SDV    &209847   &209847  &&209873  &209879  &&210106  &210115  &&210220  &210230  &&210226  &210229  &&210307 &210314   \\
6SDV    &210213   &210214  &&210240  &210245  &&210472  &210481  &&210585  &210596  &&210592  &210596  &&210673 &210680   \\
7SDV    &210326   &210327  &&210352  &210357  &&210586  &210594  &&210698  &210709  &&210710  &210714  &&210792 &210797   \\
7SDC     &210759   &210760  &&210785  &210790  &&211018  &211027  &&211131  &211142  &&211143  &211147  &&211225 &211230    \\
7SDCB  &210732   &210733  &&210756  &210761  &&210982  &210990  &&211083  &211094  &&211104  &211108  &&211177 &211182    \\
NIST&211030  &211033  &&211056  &211060  &&211287  &211295  &&211390  &211402  &&211410  &211415  &&211486 &211490    \\
\\
\multicolumn{18}{c}{Fine-structure splittings}\\
DF     &\multicolumn{2}{c}{$-4279.06$}&&\multicolumn{2}{c}{3.62} &&\multicolumn{2}{c}{7.70} &&\multicolumn{2}{c}{12.71}&&\multicolumn{2}{c}{$-409.66$} &&\multicolumn{2}{c}{5284.84}  \\
4SDV    &\multicolumn{2}{c}{$-0.45$}   &&\multicolumn{2}{c}{6.38} &&\multicolumn{2}{c}{9.28} &&\multicolumn{2}{c}{9.20} &&\multicolumn{2}{c}{3.03}    &&\multicolumn{2}{c}{7.97}    \\
5SDV    &\multicolumn{2}{c}{0.29}    &&\multicolumn{2}{c}{5.90} &&\multicolumn{2}{c}{9.06} &&\multicolumn{2}{c}{9.99} &&\multicolumn{2}{c}{3.12}    &&\multicolumn{2}{c}{6.91}    \\
6SDV    &\multicolumn{2}{c}{0.74}    &&\multicolumn{2}{c}{5.59} &&\multicolumn{2}{c}{8.90} &&\multicolumn{2}{c}{10.44}&&\multicolumn{2}{c}{3.42}    &&\multicolumn{2}{c}{6.44}    \\
7SDV    &\multicolumn{2}{c}{1.47}    &&\multicolumn{2}{c}{5.12} &&\multicolumn{2}{c}{8.36} &&\multicolumn{2}{c}{11.15}&&\multicolumn{2}{c}{4.08}    &&\multicolumn{2}{c}{5.20}     \\
7SDC     &\multicolumn{2}{c}{1.50}    &&\multicolumn{2}{c}{5.04} &&\multicolumn{2}{c}{8.24} &&\multicolumn{2}{c}{11.03}&&\multicolumn{2}{c}{4.14}     &&\multicolumn{2}{c}{5.18}     \\
7SDCB  &\multicolumn{2}{c}{1.50}    &&\multicolumn{2}{c}{4.77} &&\multicolumn{2}{c}{8.33} &&\multicolumn{2}{c}{11.10}&&\multicolumn{2}{c}{4.18}    &&\multicolumn{2}{c}{5.17}     \\
NIST&\multicolumn{2}{c}{2.86}   &&\multicolumn{2}{c}{3.98} &&\multicolumn{2}{c}{7.62} &&\multicolumn{2}{c}{12.08}&&\multicolumn{2}{c}{4.69}    &&\multicolumn{2}{c}{3.72}    \\
\enddata
\end{deluxetable}

\begin{deluxetable}{rrrrrrrrrr}
\tabletypesize{\scriptsize}
\tablecaption{\label{tab:table3}
{\footnotesize
Excitation energies (cm$^{-1}$) for states in the 2s2p$^3$ and 2p3d configurations from different computational models.
$\xi$\% is the difference between present calculations and NIST values.
}
}
\tablewidth{0pt}
\tabcolsep 5pt
\tablehead{
\colhead{States }&\colhead{DF}&\colhead{4SDV}&\colhead{5SDV}  &
\colhead{6SDV}  & \colhead{7SDV}  & \colhead{7SDC}  & \colhead{7SDCB}   & \colhead{NIST} &  $\xi$\% }
\startdata
\multicolumn{10}{c}{2s2p$^3$ }\\
2s2p$^3$    $^5$S$^o_2$   & 44604      & 44563      &  46842     &  46701     &  46912     &  46257      &  46227      & 46785   & $-1.19$   \\
2s2p$^3$    $^3$D$^o_3$   & 106226     & 99629      &  94148     &  92878     &  92842     &  92300      &  92253      & 92237   & 0.02  \\
2s2p$^3$    $^3$D$^o_2$   & 106105     & 99623      &  94134     &  92865     &  92833     &  92290      &  92260      & 92250   & 0.01  \\
2s2p$^3$    $^3$D$^o_1$   & 106027     & 99621      &  94125     &  92856     &  92827     &  92283      &  92257      & 92252   & 0.01  \\
2s2p$^3$    $^3$P$^o_2$   & 124391     & 115191     &  111596    &  109999    &  109851    &  109399     &  109366     & 109218  & 0.14  \\
2s2p$^3$    $^3$P$^o_1$   & 124253     & 115182     &  111583    &  109988    &  109844    &  109390     &  109360     & 109217  & 0.13  \\
2s2p$^3$    $^3$P$^o_0$   & 124183     & 115178     &  111576    &  109982    &  109841    &  109386     &  109365     & 109224  & 0.13  \\
2s2p$^3$    $^1$D$^o_2$   & 160740     & 150592     &  148830    &  146098    &  145719    &  144999     &  144959     & 144188  & 0.53  \\
2s2p$^3$    $^3$S$^o_1$   & 182165     & 170632     &  159546    &  157086    &  156831    &  155645     &  155609     & 155127  & 0.31  \\
2s2p$^3$    $^1$P$^o_1$   & 190938     & 180474     &  171356    &  168726    &  168315    &  167595     &  167562     & 166766  & 0.48  \\  \\
\multicolumn{10}{c}{2p3d }\\
2p3d        $^3$F$^o_2$   & 196458     & 188026     &  186817    &  186042    &  186039    &  186259     &  186235     & 186512  & $-0.15$   \\
2p3d        $^3$F$^o_3$   & 196638     & 188112     &  186898    &  186122    &  186117    &  186339     &  186303     & 186571  & $-0.14$  \\
2p3d        $^3$F$^o_4$   & 196947     & 188234     &  187006    &  186231    &  186224    &  186447     &  186395     & 186652  & $-0.14$  \\
2p3d        $^1$D$^o_2$   & 197814     & 188697     &  187428    &  186667    &  186673    &  186848     &  186807     & 187091  & $-0.15$   \\
2p3d        $^3$D$^o_1$   & 197762     & 188667     &  187575    &  186903    &  186928    &  187137     &  187103     & 187438  & $-0.18$   \\
2p3d        $^3$D$^o_2$   & 198531     & 189038     &  187610    &  186936    &  186961    &  187170     &  187130     & 187462  & $-0.18$   \\
2p3d        $^3$D$^o_3$   & 197978     & 188747     &  187651    &  186979    &  187001    &  187211     &  187165     & 187492  & $-0.17$   \\
2p3d        $^3$P$^o_2$   & 199215     & 190175     &  189069    &  188350    &  188364    &  188576     &  188538     & 188857  & $-0.17$   \\
2p3d        $^3$P$^o_1$   & 199349     & 190227     &  189121    &  188402    &  188418    &  188631     &  188586     & 188909  & $-0.17$   \\
2p3d        $^3$P$^o_0$   & 199428     & 190255     &  189149    &  188431    &  188447    &  188661     &  188612     & 188937  & $-0.17$   \\
2p3d        $^1$F$^o_3$   & 200248     & 190859     &  189631    &  188878    &  188872    &  189089     &  189047     & 189335  & $-0.15$   \\
2p3d        $^1$P$^o_1$   & 203006     & 191901     &  190594    &  189744    &  189719    &  189928     &  189887     & 190120  & $-0.12$   \\
\enddata
\end{deluxetable}

\subsection{\label{app:subsec} Transition energies, line strengths and probabilities}         

In this section we investigate the influence of electron correlation effects and the Breit interaction on the electric dipole (E1) transitions including transition energies $\triangle E$, line strengths $S$ and corresponding probabilities $A$.
In order to show these effects, the present results are presented in Table~\ref{tab:table4}
for some transitions from the 2p4f configuration as functions of the computational models.
Since the accuracy of the transition probabilities can be evaluated from the agreement between values in the Babushkin and Coulomb gauges \citep{Ekman14}, which correspond to the length and velocity gauges in the non-relativistic limit, we also present the transition rates in these two gauges.
As can be seen from Table~\ref{tab:table4}, the line strengths and the transition rates of the strong lines are well converged in both gauges.
Moreover, the consistency of the line strengths and transition rates in the two gauges are quite good in the 7SDCB model.
In comparison with experimental \textbf{value}~\citep{Mar00}, good agreement is found as well.
For most of the weak lines, however, we observed that good convergence merely appear in the Babushkin (length) gauge but not in the Coulomb (velocity) gauge. Actually, it is indeed difficult to converge transition rates in the Coulomb gauge for the weak lines, since the transition operator in the Coulomb gauge is more sensitive to electron correlations than that in the Babushkin gauge.  
For this reason, we recommend the transition rates in the Babushkin (length) gauge to be used in astrophysical applications.
%

%
The uncertainties of the transition rates in the Babushkin (length) gauges are estimated based on the convergence trends.
It is seen that the values change by about 5\% from the 6SDV model to 7SDV, except for some weak lines, for example, in the 2p4f - 2p4d and 2p4f - 2s2p$^3$ transition arrays. For the former lines the small transition energies are associated with large relative uncertainties that lead to poor convergence for the transition rates that have uncertainties reaching 10\%. However, these uncertainties can be reduced by rescaling the transition rates with experimental energies as we will show later. The 2p4f - 2s2p$^3$ transition is a two-electron-one-photon process and thus sensitive to electron correlation effects \citep{two}. In the present calculation, the uncertainty for these transitions is about 10\% - 15\%.

%


\begin{deluxetable}{lcccccccccccccccccccc}
\tabletypesize{\scriptsize}
\tablecaption{\label{tab:table4}
{\footnotesize
Line strengths $S$ (in a.u.) and probabilities $A$ (in s$^{-1}$) of E1 transitions involving 2p4f and lower configurations together with the corresponding transition energies $\Delta$$E$ in (cm$^{-1}$).
%
%
The number in the square bracket represents the power of 10.
B(len) and C(vel) denote values in Babushkin and Coulomb gauges, respectively.
Exp. are the experimental \textbf{value} taken from Ref.~\citep{Mar00}.
}}
\tablewidth{0pt}
\tabcolsep 2.5pt
\tablehead{
& &\multicolumn{2}{c}{ $S$}&&\multicolumn{2}{c}{$A$} &&& \multicolumn{2}{c}{ $S$}&&\multicolumn{2}{c}{$A$} &&& \multicolumn{2}{c}{ $S$}&&\multicolumn{2}{c}{$A$}\\
\cline{3-4}  \cline{6-7}  \cline{10-11} \cline{13-14} \cline{17-18} \cline{20-21}
\colhead{Model} & \colhead{ $\Delta$$E$ } & \colhead{B(len) } & \colhead{C(vel) } &   & \colhead{B(len)}&\colhead{C(vel)}& &
                  \colhead{ $\Delta$$E$ } & \colhead{B(len) } & \colhead{C(vel) } &   & \colhead{B(len)}&\colhead{C(vel)}& &
                  \colhead{ $\Delta$$E$ } & \colhead{B(len) } & \colhead{C(vel) } &   & \colhead{B(len)}&\colhead{C(vel)}  }
\startdata
\multicolumn{21}{c}{2p4f -- 2p3d}	\\																																				&	\multicolumn{6}{c}{ G(7/2)$_3$ -- $^3$F$^o_2$}&	&	\multicolumn{6}{c}{D(5/2)$_2$ -- $^1$P$^o_1$}	&	&	\multicolumn{6}{c}{F(7/2)$_3$ -- $^3$P$^o_2$}	\\	
DF	&	22879	&	3.70[1]	&	4.14[1]	&	&	1.28[8]	&	1.44[8]	&	&	16058 	&	1.39[0]	&	1.30[0]	&	&	2.33[6]	&	2.18[6]	&	&	19874 	&	3.24[-1]	&	3.38[-1]	&	&	7.35[5]	&	7.68[5]	\\	
4SDV	&	17969	&	3.91[1]	&	6.80[1]	&	&	6.57[7]	&	1.14[8]	&	&	14222 	&	2.25[1]	&	4.15[1]	&	&	2.63[7]	&	4.84[7]	&	&	15581 	&	3.72[-2]	&	7.06[-2]	&	&	4.07[4]	&	7.73[4]	\\	
5SDV	&	23288	&	3.59[1]	&	4.03[1]	&	&	1.31[8]	&	1.47[8]	&	&	19636 	&	2.22[1]	&	2.42[1]	&	&	6.80[7]	&	7.44[7]	&	&	20803 	&	1.62[-2]	&	1.91[-2]	&	&	4.23[4]	&	4.98[4]	\\	
6SDV	&	24430	&	3.59[1]	&	3.72[1]	&	&	1.52[8]	&	1.57[8]	&	&	20852 	&	2.17[1]	&	2.18[1]	&	&	7.97[7]	&	8.00[7]	&	&	21890 	&	1.30[-2]	&	1.40[-2]	&	&	3.96[4]	&	4.24[4]	\\	
7SDV	&	24547	&	3.59[1]	&	3.59[1]	&	&	1.54[8]	&	1.54[8]	&	&	20995 	&	2.17[1]	&	2.10[1]	&	&	8.15[7]	&	7.87[7]	&	&	21989 	&	1.27[-2]	&	1.26[-2]	&	&	3.91[4]	&	3.87[4]	\\	
7SDC	&	24760	&	3.56[1]	&	3.50[1]	&	&	1.57[8]	&	1.54[8]	&	&	21219 	&	2.18[1]	&	2.06[1]	&	&	8.43[7]	&	7.99[7]	&	&	22209 	&	1.13[-2]	&	1.10[-2]	&	&	3.58[4]	&	3.49[4]	\\	
7SDCB	&	24747	&	3.93[1]	&	3.85[1]	&	&	1.72[8]	&	1.69[8]	&	&	21222 	&	2.20[1]	&	2.08[1]	&	&	8.51[7]	&	8.06[7]	&	&	22218 	&	1.17[-2]	&	1.13[-2]	&	&	3.70[4]	&	3.59[4]	\\	
NIST	&	24776	&		&		&	&		&		&	&	21295 	&		&		&	&		&		&	&	22199 	&		&		&	&		&		\\	
Exp.	&		&		&		&	&	\multicolumn{2}{c}{ 1.30[8]}			&	&		&		&		&	&		&		&	&		&		&		&	&		&		\\	\\
\multicolumn{21}{c}{2p4f -- 2p4d}																																				\\		&	\multicolumn{6}{c}{G(9/2)$_5$ -- $^3$F$^o_4$}				 						&	&	\multicolumn{6}{c}{G(9/2)$_4$ -- $^3$F$^o_3$}										&	&	\multicolumn{6}{c}{F(7/2)$_4$ -- $^3$D$^o_3$}		 		  						\\	
5SDV	&	263 	&	3.24[2]	&	8.26[3]	&	&	1.08[3]	&	2.76[4]	&	&	379 	&	7.68[1]	&	1.10[3]	&	&	9.45[2]	&	1.35[4]	&	&		&		&		&	&		&		\\	
6SDV	&	1274 	&	3.26[2]	&	3.85[2]	&	&	1.24[5]	&	1.47[5]	&	&	1392 	&	7.65[1]	&	8.83[1]	&	&	4.65[4]	&	5.36[4]	&	&	535 	&	1.73[2]	&	5.58[2]	&	&	5.96[3]	&	1.92[4]	\\	
7SDV	&	1330 	&	3.25[2]	&	2.66[2]	&	&	1.41[5]	&	1.15[5]	&	&	1449 	&	7.65[1]	&	6.07[1]	&	&	5.25[4]	&	4.16[4]	&	&	580 	&	1.73[2]	&	2.94[2]	&	&	7.64[3]	&	1.29[4]	\\	
7SDV	&	1552 	&	3.25[2]	&	1.92[2]	&	&	2.24[5]	&	1.32[5]	&	&	1673 	&	7.64[1]	&	4.50[1]	&	&	8.05[4]	&	4.74[4]	&	&	804 	&	1.73[2]	&	1.49[2]	&	&	2.02[4]	&	1.74[4]	\\	
7SDCB	&	1555 	&	3.25[2]	&	1.91[2]	&	&	2.25[5]	&	1.32[5]	&	&	1658 	&	7.62[1]	&	4.46[1]	&	&	7.82[4]	&	4.57[4]	&	&	821 	&	1.87[2]	&	1.59[2]	&	&	2.33[4]	&	1.97[4]	\\	
NIST	&	1566 	&		&		&	&	 	&	 	&	&	1664 	&		&		&	&		&		&	&	759 	&		&		&	&		&		\\	\\
\multicolumn{21}{c}{2p4f -- 2s2p$^3$}												 	  																							\\		&	\multicolumn{6}{c}{D(3/2)$_2$  --  $^1$P$^o_1$}										&	&	\multicolumn{6}{c}{D(5/2)$_2$ -- $^1$P$^o_1$}				 						&	&	\multicolumn{6}{c}{D(3/2)$_2$  -- $^3$P$^o_2$ }										\\		
DF	&	28545 	&	2.56[0]	&	3.22[0]	&	&	2.41[7]	&	3.03[7]	&	&	28126 	&	3.46[-1]	&	4.08[-1]	&	&	3.12[6]	&	3.68[6]	&	&	95092	&	1.50[-4]	&	1.64[-3]	&	&	5.21[4]	&	5.72[5]	\\	
4SDV	&	25736 	&	7.08[-1]	&	1.24[0]	&	&	4.89[6]	&	8.57[6]	&	&	25649 	&	7.73[-1]	&	1.36[0]	&	&	5.28[6]	&	9.27[6]	&	&	91019	&	4.11[-4]	&	3.25[-3]	&	&	1.26[5]	&	9.93[5]	\\	
5SDV	&	38958 	&	1.72[-1]	&	2.38[-1]	&	&	4.13[6]	&	5.70[6]	&	&	38873 	&	1.88[-1]	&	2.59[-1]	&	&	4.47[6]	&	6.16[6]	&	&	98718	&	8.68[-4]	&	2.13[-3]	&	&	3.38[5]	&	8.30[5]	\\	
6SDV	&	41954 	&	1.28[-1]	&	1.67[-1]	&	&	3.84[6]	&	5.01[6]	&	&	41870 	&	1.41[-1]	&	1.83[-1]	&	&	4.20[6]	&	5.45[6]	&	&	100681	&	1.13[-3]	&	1.94[-3]	&	&	4.67[5]	&	8.03[5]	\\	
7SDV	&	42482 	&	1.19[-1]	&	1.45[-1]	&	&	3.70[6]	&	4.51[6]	&	&	42399 	&	1.35[-1]	&	1.63[-1]	&	&	4.16[6]	&	5.02[6]	&	&	100946	&	1.28[-3]	&	1.82[-3]	&	&	5.32[5]	&	7.58[5]	\\	
7SDC	&	43635 	&	1.06[-1]	&	1.30[-1]	&	&	3.58[6]	&	4.37[6]	&	&	43552 	&	1.20[-1]	&	1.45[-1]	&	&	4.02[6]	&	4.86[6]	&	&	101831	&	1.20[-3]	&	1.73[-3]	&	&	5.12[5]	&	7.39[5]	\\	
7SDCB	&	43621 	&	1.08[-1]	&	1.32[-1]	&	&	3.64[6]	&	4.45[6]	&	&	43547 	&	1.21[-1]	&	1.46[-1]	&	&	4.04[6]	&	4.89[6]	&	&	101817	&	1.18[-3]	&	1.70[-3]	&	&	5.04[5]	&	7.28[5]	\\	
NIST	&	44725 	&		&		&	&		&		&	&	44650 	&		&	 	&	&	 	&		&	&	102273 	&		&		&	&		&		\\	\\
\multicolumn{21}{c}{2p4f -- 2p3s}																																				\\		&	\multicolumn{6}{c}{D(5/2)$_3$ -- $^3$P$^o_2$ }										&	&	\multicolumn{6}{c}{D(3/2)$_2$ -- $^3$P$^o_2$}										&	&	\multicolumn{6}{c}{F(5/2)$_2$ -- $^3$P$^o_1$ }										\\		
DF	&	46955 	&	2.25[-2]	&	4.26[-2]	&	&	6.74[5]	&	1.28[6]	&	&	46964	&	4.36[-2]	&	9.40[-2]	&	&	1.83[6]	&	3.95[6]	&	&	54451	&	1.73[-2]	&	1.45[-2]	&	&	1.13[6]	&	9.50[5]	\\	
4SDV	&	44499 	&	1.63[-2]	&	3.83[-2]	&	&	4.15[5]	&	9.77[5]	&	&	44589	&	2.50[-2]	&	6.81[-2]	&	&	8.99[5]	&	2.45[6]	&	&	55280	&	1.43[-4]	&	4.29[-4]	&	&	9.80[3]	&	2.94[4]	\\	
5SDV	&	60610 	&	9.66[-3]	&	1.96[-2]	&	&	6.22[5]	&	1.27[6]	&	&	60698	&	1.03[-3]	&	2.10[-3]	&	&	9.33[4]	&	1.90[5]	&	&	60388	&	1.51[-4]	&	2.82[-4]	&	&	1.35[4]	&	2.51[4]	\\	
6SDV	&	61908 	&	8.90[-3]	&	1.62[-2]	&	&	6.11[5]	&	1.11[6]	&	&	61995	&	9.28[-4]	&	1.69[-3]	&	&	8.96[4]	&	1.63[5]	&	&	61684	&	1.43[-4]	&	2.47[-4]	&	&	1.36[4]	&	2.35[4]	\\	
7SDV	&	62037 	&	9.57[-3]	&	1.52[-2]	&	&	6.61[5]	&	1.05[6]	&	&	62124	&	1.00[-3]	&	1.59[-3]	&	&	9.75[4]	&	1.54[5]	&	&	61808	&	1.49[-4]	&	2.27[-4]	&	&	1.43[4]	&	2.17[4]	\\	
7SDC	&	62251 	&	9.31[-3]	&	1.49[-2]	&	&	6.50[5]	&	1.04[6]	&	&	62339	&	9.72[-4]	&	1.55[-3]	&	&	9.54[4]	&	1.52[5]	&	&	62029	&	1.44[-4]	&	2.18[-4]	&	&	1.39[4]	&	2.11[4]	\\	
7SDCB&	62259 	&	9.49[-3]	&	1.52[-2]	&	&	6.63[5]	&	1.06[6]	&	&	62337	&	9.58[-4]	&	1.53[-3]	&	&	9.40[4]	&	1.50[5]	&	&	62023	&	1.09[-4]	&	1.67[-4]	&	&	1.06[4]	&	1.62[4]	\\	
NIST	&	62334 	&		&		&	&		&		&	&	62414 	&		&		&	&		&		&	&	62093 	&		&		&	&		&		\\	\\
\multicolumn{21}{c}{2p4f -- 2p4s}																																				\\		&	\multicolumn{6}{c}{ D(5/2)$_2$ -- $^1$P$^o_1$ }										&	&	\multicolumn{6}{c}{ D(3/2)$_2$ -- $^3$P$^o_1$}										&	&	\multicolumn{6}{c}{D(3/2)$_1$  -- $^3$P$^o_2$ }										\\		
DF	&	10727 	&	5.71[-1]	&	4.71[-1]	&	&	2.86[5]	&	2.35[5]	&	&	13181	&	3.76[-2]	&	2.27[-2]	&	&	3.49[4]	&	2.11[4]	&	&	7451 	&	1.48[1]	&	1.94[1]	&	&	4.15[6]	&	5.43[6]	\\	
4SDV	&	6899 	&	2.20[0]	&	5.98[0]	&	&	2.93[5]	&	7.95[5]	&	&	8859	&	4.67[-1]	&	1.06[0]	&	&	1.32[5]	&	2.98[5]	&	&	8713 	&	1.48[-2]	&	3.38[-2]	&	&	6.59[3]	&	1.51[4]	\\	
5SDV	&	12046 	&	1.53[0]	&	1.57[0]	&	&	1.08[6]	&	1.11[6]	&	&	\textbf{13543}	&	\textbf{4.29[-1]}	&	\textbf{4.37[-1]}	&	&	\textbf{4.32[5]}	&	\textbf{4.40[5]}	&	&	13400 	&	1.33[-2]	&	1.36[-2]	&	&	2.17[4]	&	2.21[4]	\\	
6SDV	&	13172 	&	1.23[0]	&	1.10[0]	&	&	1.14[6]	&	1.02[6]	&	&	14605	&	3.74[-1]	&	3.34[-1]	&	&	4.72[5]	&	4.22[5]	&	&	14459 	&	1.15[-2]	&	1.03[-2]	&	&	2.35[4]	&	2.11[4]	\\	
7SDV	&	13297 	&	1.12[0]	&	9.63[-1]	&	&	1.07[6]	&	9.18[5]	&	&	14716	&	3.60[-1]	&	3.08[-1]	&	&	4.65[5]	&	3.98[5]	&	&	14570 	&	1.11[-2]	&	9.51[-3]	&	&	2.33[4]	&	1.99[4]	\\	
7SDC	&	13552 	&	1.11[0]	&	9.19[-1]	&	&	1.12[6]	&	9.27[5]	&	&	14935	&	3.63[-1]	&	3.01[-1]	&	&	4.89[5]	&	4.06[5]	&	&	14786 	&	1.11[-2]	&	9.20[-3]	&	&	2.42[4]	&	2.01[4]	\\	
7SDCB&	13554 	&	1.11[0]	&	9.18[-1]	&	&	1.12[6]	&	9.26[5]	&	&	14914	&	3.46[-1]	&	2.87[-1]	&	&	4.65[5]	&	3.86[5]	&	&	14787 	&	1.11[-2]	&	9.22[-3]	&	&	2.43[4]	&	2.01[4]	\\	
NIST	&	13556 	&		&		&	&		&		&	&	14898 	&		&		&	&		&		&	&	14775 	&		&		&	&		&		\\	\\
\enddata
\end{deluxetable}
\clearpage

\subsection{\label{app:subsec} Evaluations of $gf$ for terms of the 2p4f configuration}           

Oscillator strengths for terms belonging to the 2p4f configuration were provided  by Kelly and Wiese~\citep{Kelly64} and the TOPbase of Opacity Project (\textit{OP}) data~\citep{OP95}.
In order to evaluate the compiled data, we make comparisons with the present values.
One should keep in mind that the previous calculations were non-relativistic and based on the LS-coupling scheme.
Without loss of generality, we list the $gf$ values for transitions from the 2p4f to the 2p3d configuration in Table~\ref{tab:table5}.
It can be seen from this table that our calculations are consistent with other results.
The small discrepancies, however, are indicators of the neglected relativistic effects in previous calculations.
The importance of the relativistic effects can be seen more clearly in term separations that mainly result from the spin-orbital interaction of the 2p electron.
Using the excitation energies reported in Table~\ref{tab:table2}, we obtain the term separations as the difference between the weighted average energies over the pair of levels.
The values are listed in Table~\ref{tab:table6}.
For comparison, we also show the results obtained with NIST values.
It is found that present calculations are in excellent agreement with NIST the values, but differ remarkably from the ones of the Opacity Project due to neglect of relativistic effects and inadequate consideration of electron correlations in the latter.
%
This means that non-relativistic calculations and the associated LS-coupling scheme are inappropriate for the case under investigation.
%



\begin{deluxetable}{rrrrrr}
\tabletypesize{\scriptsize}
\tablecaption{\label{tab:table5}
{\footnotesize
The comparisons of the term $gf$ for the 2p4f -- 2p3d transitions.
VE, HFS, and OP are values taken from~\citep{Victor88}, ~\citep{Kelly64,Wiese65}, and~\citep{OP95}.
%
%
}}
\tablewidth{0pt}
\tabcolsep 7pt
\tablehead{
 & &\multicolumn{4}{c}{ $gf$ }    \\
\cline{3-6}
    \colhead{2p4f -- 2p3d} && \colhead{This work} & \colhead{VE}&
\colhead{OP}&\colhead{ HFS}
}
\startdata
          F-D$^o$      &&14.16  &15.64  &16.35  &16.15           \\
         F-F$^o$       && 3.33  & 1.90 &2.02   &           \\   \\
         G-F$^o$       && 21.07 &22.27 &22.98  &       \\  \\
         D-P$^o$       &&11.08  &11.46 &11.32  &10.89         \\
         D-D$^o$       && 2.14  &1.99  &2.01    &            \\
         D-F$^o$       &&0.10   &0.06  &0.05   &            \\
\enddata
\end{deluxetable}


\begin{deluxetable}{rrrr}
\tabletypesize{\scriptsize}
\tablecaption{\label{tab:table6}
{\footnotesize
The separations (in cm$^{-1}$) in F, G and D terms of the 2p4f configuration.
OP are values obtained with the Opacity Project data~\citep{OP95}.
}
}
\tablewidth{0pt}
\tabcolsep 10pt
\tablehead{
& \multicolumn{3}{c}{ Term Splitting (cm$^{-1}$)  }     \\
\cline{2-4}

\colhead{ Array }& \colhead{F}& \colhead{G} & \colhead{D} }
\startdata
This work	&	26.34 	&	101.03 	&	74.43 \\
OP	        &	7.68 	&	26.34 	&	15.36 \\
NIST	    &	27.24 	&	103.87 	&	76.48 \\
\enddata
\end{deluxetable}

\subsection{\label{app:subsec} Atomic parameters of the 2p4f configuration }    

\begin{figure}[!th]
\centering
\includegraphics[width=13cm]{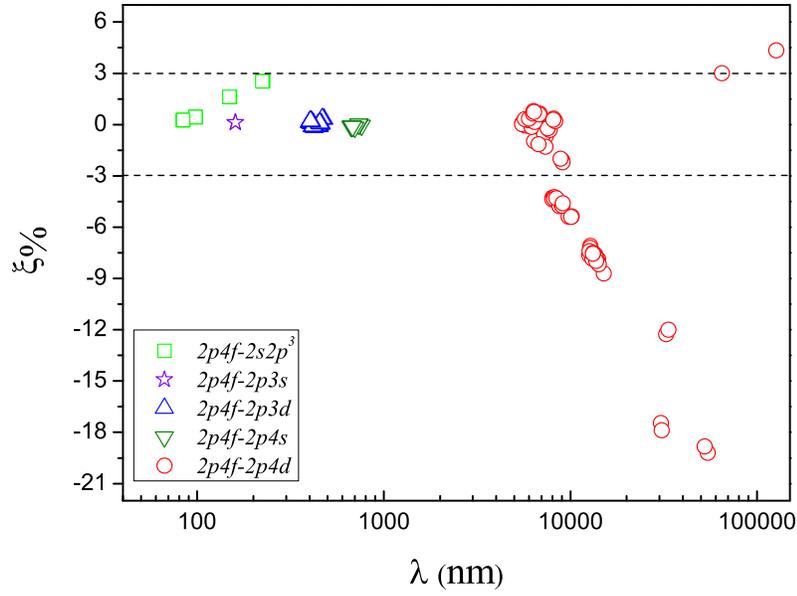}
\caption{\label{fig:Fig2}
The relative difference ($\xi\%$) in wavelengths between present calculations and NIST values for all transitions from the 2p4f configuration.
 }
\end{figure}

Wavelengths $\lambda$, weighted oscillator strengths $gf$ and transition probabilities $A$ of E1 transitions from levels in the 2p4f configuration to all lower-lying levels in N$^{+}$ are reported in Table~\ref{tab:table7}.
These data are arranged according to different transition arrays like 2p4f -- 2s2p$^3$, 2p4f -- 2p3s, 2p4f -- 2p3d, and so on.
In the present work, we only present results associated with $gf$ values larger than $5 \times 10^{-4}$ in the Babushkin (length) gauge.
The relative difference in wavelengths ($\xi\%$) between the present calculation and NIST values is listed in the fifth column of Table~\ref{tab:table7}. \textbf{For convenience, this was also illustrated in Figure~\ref{fig:Fig2}.}
It can be seen that the difference is about $0.2\%$ for the 2p4f -- 2p3s, 2p4f -- 2p3d and 2p4f -- 2p4s transitions.
Some transitions down to 2s2p$^3$, e.g. $^1$D$^o_2$ and $^1$P$^o_1$, are off by 1.6\% -- 2.5\%.
%
%
It should be noted that the transitions between states of 2p4f and 2p4d configurations are exceptions.
The transition energies are small and thus very hard to obtain accurately as they result from the subtraction of two equally large numbers
%

Transition rates in Babushkin (length) gauge are presented in the 7th column of Table~\ref{tab:table7}.
The available experimental transition rates for the transition 2p4f -- 2p3d are also displayed for comparison.
It can be shown that present calculations are in reasonable agreement with the measurements by Mar et al.
The only large discrepancy is found for the transition from 4f F(5/2)$_2$ to 3d $^3$F$^o_2$.
%
%

It should be pointed out that the errors in the wavelengths lead to errors in the calculated transition rates, especially for the transitions with large wavelengths, e.g. the 2p4f -- 2p4d transitions.
The errors in the transitions, however, can be corrected by scaling the rates with experimental wavelengths.
We should stressed that these lines are hardly observed in experiments due to small branching ratios.
Even though they are of little diagnostic importance we still present scaled $gf$ values in Table~\ref{tab:table8} for lines where the difference in wavelength compared to NIST is larger than $3\%$.
The final scaled results are obviously improved.
%

%
Liu pointed out that $\lambda$404.1 is the strongest line among the ones from the 2p4f configuration~\citep{Liu00}.
This is confirmed by our calculations. Moreover, we found that the $gf$ value of the line with $\lambda=424.1$ nm is large.
This may be the reason why there is much work focusing on these two lines~\citep{Escalante91,Liu00,Fang11}.
In addition, we found that in the infrared region there is a strong line produced by the transition from 4f G(9/2)$_5$ to 4d $^3$F$^o_4$ with $gf(=1.54)$.
%

%
With regard to plasma diagnostics, accurate atomic data are indispensable.
For example, Prueitt used a group of multiplet lines with $\lambda$403.51nm, $\lambda$404.13 and $\lambda$404.35, namely the transition between 2p4f$~^3$G and 2p3d$~^3$F$^o$, to determine the temperature of plasmas produced by lightnings~\citep{Prueitt63}.
The values used to diagnose the plasma in that work deviate substantially from the present results.
With respect to the accuracy of present calculations, some analysis based on old atomic data should be re-made.


\begin{deluxetable}{rrrrrrrrr}
\tabletypesize{\scriptsize}
\tablecaption{\label{tab:table7}
{\footnotesize
Wavelengths $\lambda$, weighted oscillator strengths $gf$ and transition
probabilities $A$ of E1 transitions from the 2p4f configuration.
Obs. are taken from NIST except for those with superscript.
$^{a,b,c}$ are referred to Ref.~\citep{Eriksson83},~\citep{Mar00}, and~\citep{Marquette00}.
%
%
$\sigma$$^b$ are the uncertainties of experimental rates~\citep{Mar00}.
The number in the square bracket represents the power of 10.
}}
\tablewidth{0pt} \tabcolsep 5pt
\tablehead{
&&\multicolumn{3}{c}{ $\lambda$(nm) }&& \multicolumn{3}{c}{ $A$ ( s$^{-1}$)}  \\
\cline{3-5} \cline{7-9}
\colhead{Upper} &\colhead{Lower} &\colhead{Calc.}&\colhead{Obs.}&\colhead{$\xi$\%} &  \colhead{$gf$}
&  \colhead{Calc.} & \colhead{ Exp\tablenotemark{b}} & \colhead{$\sigma$\tablenotemark{b}}
 }
\startdata
\multicolumn{9}{c}{2p4f -- 2s2p$^3$}\\
2p4f	D(3/2)$_1$	& 2s2p$^3$	$^3$P$^o_0$   &	98.220 	    &   97.787 	    &	0.44 	  	& 9.23[-4]		&	2.13[6] &		    &				\\	
2p4f	D(3/2)$_1$	& 2s2p$^3$	$^3$P$^o_1$   &	98.216 	    &   97.780 	    &	0.45 	 	& 6.93[-4]		&	1.60[6] &		    &				\\
2p4f	F(5/2)$_2$	& 2s2p$^3$	$^3$D$^o_1$   &	84.405    	&   84.188 	    &	0.26 	 	& 2.63[-3]		&	4.93[6] & 		    &				\\	
2p4f	D(5/2)$_2$	& 2s2p$^3$	$^3$P$^o_1$   &	98.282 	    &   97.849 	    &	0.44 	  	& 9.23[-4]		&	1.27[6] &		    &				\\	
2p4f	D(3/2)$_2$	& 2s2p$^3$	$^3$P$^o_1$   &	98.211 	    &   97.777 	    &	0.44 	 	& 1.10[-3]		&	1.52[6] &		    &				\\	
2p4f	F(5/2)$_2$	& 2s2p$^3$	$^1$P$^o_1$   &	231.634     &   225.901     &	2.54 	 	& 1.02[-3]		&	2.53[5] & 		    &			    \\	
2p4f	D(5/2)$_2$	& 2s2p$^3$	$^1$P$^o_1$   &	229.638     &	223.967     &	2.53 	  	& 1.60[-2]		&	4.04[6] &		    &				\\	
2p4f	D(3/2)$_2$	& 2s2p$^3$	$^1$P$^o_1$   &	229.249     &	223.590     &	2.53 	 	& 1.44[-2]		&	3.64[6] &		    &				\\	
2p4f	D(5/2)$_2$	& 2s2p$^3$	$^1$D$^o_2$   &	151.173     &	148.749     &	1.63 	 	& 1.79[-3]		&	1.05[6] &		    &				\\	
2p4f	D(3/2)$_2$	& 2s2p$^3$	$^1$D$^o_2$   &	151.004     &	148.583	    &	1.63 	 	& 1.61[-3]		&	9.42[5] &		    &				\\	
2p4f	F(5/2)$_3$	& 2s2p$^3$	$^3$D$^o_2$   &	84.409 	    &	84.189 	    &	0.26 	 	& 1.89[-3]		&	2.52[6] &		    &				\\	
2p4f	F(7/2)$_3$	& 2s2p$^3$	$^3$D$^o_2$   &	84.391 	    &	84.171 	    &	0.26 	 	& 1.85[-3]		&	2.48[6] &		    &				\\	
2p4f	D(5/2)$_3$	& 2s2p$^3$	$^3$P$^o_2$   &	98.291 	    &	97.854 	    &	0.45 	 	& 3.62[-3]		&	3.57[6] &		    &				\\	
2p4f	F(5/2)$_3$	& 2s2p$^3$	$^1$D$^o_2$   &	152.039     &	149.606     &	1.63 	    & 1.66[-2]		&	6.86[6] &		    &				\\	
2p4f	F(7/2)$_3$	& 2s2p$^3$	$^1$D$^o_2$   &	151.982     &	149.548     &	1.63 	    & 1.40[-2]		&	5.78[6] &		    &				\\	
2p4f	G(7/2)$_3$	& 2s2p$^3$	$^1$D$^o_2$   &	151.463     &	149.033     &	1.63 	    & 2.20[-3]		&	9.12[5] &		    &				\\	
2p4f	D(5/2)$_3$	& 2s2p$^3$	$^1$D$^o_2$   &	151.182     &	148.760     &	1.63 	    & 7.43[-4]		&	3.10[5] &		    &				\\
2p4f	D(5/2)$_3$	& 2s2p$^3$	$^3$D$^o_3$   &	84.139 	    &	83.911 	    &	0.27 	 	& 7.76[-4]		&	1.04[6] &		    &				\\
2p4f	F(7/2)$_4$	& 2s2p$^3$	$^3$D$^o_3$   &	84.382 	    &	84.159 	    &	0.27 	 	& 4.88[-3]		&	5.08[6] &		    &				\\	
2p4f	G(7/2)$_4$	& 2s2p$^3$	$^3$D$^o_3$   &	84.219 	    &	83.993 	    &	0.27 	 	& 6.26[-4]		&	6.54[5] &		    &				\\ \\
\multicolumn{9}{c}{2p4f -- 2p3s}\\
2p4f	D(3/2)$_2$	&	2p3s	$^3$P$^o_1$	&	160.071 	&   159.872     &	0.12 	  	& 5.04[-4]		&	2.62[5] &		    &				\\
2p4f	D(5/2)$_3$	&	2p3s	$^3$P$^o_2$	&	160.618 	&	160.426     &	0.12 	  	& 1.79[-3]		&	6.63[5] &		    &			  \\ \\
\multicolumn{9}{c}{2p4f -- 2p3d}\\
2p4f	D(3/2)$_1$	&	2p3d	$^3$P$^o_0$	&	443.157 	&  443.472      &	$-0.07$ 	 	& 9.30[-1]		&	1.05[8] &		    &			\\	
2p4f	D(3/2)$_1$	&	2p3d	$^3$D$^o_1$	&	415.378 	&  415.817      &	$-0.11$ 		& 1.96[-1]		&	2.53[7] &		    &			\\	
2p4f	D(3/2)$_1$	&	2p3d	$^3$P$^o_1$	&	442.645 	&  442.921      &	$-0.06$ 		& 7.23[-1]		&	8.21[7] &		    &			\\	
2p4f	F(5/2)$_2$	&	2p3d	$^3$D$^o_1$	&	423.182 	&  423.812      &	$-0.15$ 	 	& 2.44[0] 		&	1.81[8] &		    &				\\	
2p4f	D(5/2)$_2$	&	2p3d	$^3$D$^o_1$	&	416.568 	&  417.056      &	$-0.12$ 	  	& 1.18[-2]		&	9.07[5] &		    &				\\	
2p4f	D(3/2)$_2$	&	2p3d	$^3$D$^o_1$	&	415.289 	&  415.753      &	$-0.11$ 	 	& 5.03[-2]		&	3.89[6] &		    &				\\	
2p4f	F(5/2)$_2$	&	2p3d	$^3$P$^o_1$	&	451.517 	&  452.003      &	$-0.11$ 	 	& 2.83[-2]		&	1.85[6] &		    &				\\	
2p4f	D(5/2)$_2$	&	2p3d	$^3$P$^o_1$	&   443.996     &  444.326      &	$-0.07$ 	 	& 1.01[0]		&	6.83[7] & 6.95[7]   & 16\%			\\	
                    &                       &               &  444.20$^b$   &                       &              &          &           &              \\
2p4f	D(3/2)$_2$	&	2p3d	$^3$P$^o_1$	&   442.544     &  442.848      &	$-0.07$ 	 	& 1.03[0]		&	7.01[7] & 5.68[7]   & 50\% 			\\
                    &                       &               &  442.72$^b$   &                	&              &          &           &              \\ 
2p4f	F(5/2)$_2$	&	2p3d	$^1$P$^o_1$	&	479.694 	&	478.179     &	0.32 	  	 	& 6.55[-2]		&	3.80[6] &		    &				\\
                    &                       &               & 478.043$^a$   &                	&              &          &           &              \\ 
2p4f	D(5/2)$_2$	&	2p3d	$^1$P$^o_1$	&	471.214 	&	469.596     &	0.34 	     	& 1.42[0]   	&	8.51[7] & 6.07[7]   & 12\%  		\\
                    &                       &               &  469.46$^b$   &                	&               &          &           &                \\ 
2p4f	D(3/2)$_2$	&	2p3d	$^1$P$^o_1$	&	469.577 	&	467.944     &	0.35 	 	 	& 1.45[0]		&	8.77[7] &		    &				\\	
2p4f	D(3/2)$_1$	&	2p3d	$^3$F$^o_2$	&	400.920 	&	400.400     &	0.13 	 	 	& 8.64[-3]		&	1.20[6] &		    &				\\	
2p4f	D(3/2)$_1$	&	2p3d	$^3$D$^o_2$	&	415.846 	&	416.233     &	$-0.09$ 	  	& 7.15[-2]		&	9.19[6] &		    &				\\	
2p4f	D(3/2)$_1$	&	2p3d	$^3$P$^o_2$	&	441.720 	&	441.907     &	$-0.04$ 	 	& 5.19[-2]		&	5.92[6] &		    &				\\	
2p4f	F(5/2)$_2$	&	2p3d	$^3$F$^o_2$	&	408.185     &	407.808     &	0.09 	   		& 2.64[-1]		&	2.12[7] & 8.00[6]   & 42\%  		\\	
                    &                       &               &  407.69$^b$   &                      &               &          &           &               \\ 
2p4f	F(5/2)$_2$	&	2p3d	$^1$D$^o_2$	&	417.949 	&	417.684     &	0.06 	   		& 1.39[-2]		&	1.06[6] &		    &				\\	
2p4f	D(5/2)$_2$	&	2p3d	$^1$D$^o_2$	&	411.497 	&	411.120     &	0.09 	 	 	& 2.24[-1]		&	1.76[7] &		    &			\\	
2p4f	D(3/2)$_2$	&	2p3d	$^1$D$^o_2$	&	410.249 	&	409.854     &	0.10 		 	& 2.30[-1]		&	1.82[7] &		    &			\\	
2p4f	F(5/2)$_2$	&	2p3d	$^3$D$^o_2$	&	423.667 	&	424.243 	&	$-0.14$ 	 	& 3.42[-1]		&	2.54[7] &		    &				\\	
2p4f	D(5/2)$_2$	&	2p3d	$^3$D$^o_2$	&	417.038 	&	417.474 	&	$-0.10$   	 	& 3.09[-1]  	&	2.37[7] & 1.20[7]   & 30\%  	\\	
                    &                       &               &   417.36$^b$  &                       &               &          &           &                \\
2p4f	D(3/2)$_2$	&	2p3d	$^3$D$^o_2$	&	415.756 	&	416.168 	&	$-0.10$ 	    & 1.27[-1]		&	9.77[6] &		    &			  \\	
2p4f	F(5/2)$_2$	&	2p3d	$^3$P$^o_2$	&	450.555 	&	450.947 	&	$-0.09$ 	 	& 8.59[-3]		&	5.65[5] &		    &				\\	
2p4f	D(5/2)$_2$	&	2p3d	$^3$P$^o_2$	&	443.065 	&	443.306 	&	$-0.05$ 	 	& 3.18[-1]		&	2.16[7] &		    &			   \\	
2p4f	D(3/2)$_2$	&	2p3d	$^3$P$^o_2$	&	441.619 	&	441.834 	&	$-0.05$    	 	& 4.16[-1]		&	2.85[7] & 2.33[7]   & 14\%  	   \\
                    &                       &               &    441.71$^b$ &                      &               &          &           &              \\ 
2p4f	F(7/2)$_3$	&	2p3d	$^3$F$^o_2$	&	407.805 	&	407.420 	&	0.09      	 	& 9.66[-1]  	&	5.54[7] & 4.99[7]   & 19\%     	  \\
                    &                       &               &   407.30$^b$  &                      &               &          &           &              \\ 
2p4f	G(7/2)$_3$	&	2p3d	$^3$F$^o_2$	&	404.085 	&	403.622 	&	0.11 	    	 & 2.95[0]		& 1.72[8]  & 1.30[8]   &	7\%			  \\	
                    &                       &               &   403.51$^b$  &                      &               &          &           &                \\ 
2p4f	F(5/2)$_3$	&	2p3d	$^1$D$^o_2$	&	417.975 	&	417.734 	&	0.06 	   		& 2.22[0]   	& 1.21[8]  & 1.13[8]   & 19\%  		  \\	
                    &                       &               &   417.62$^b$  &                       &               &          &           &                \\ 
2p4f	F(7/2)$_3$	&	2p3d	$^1$D$^o_2$	&	417.552 	&	417.277 	&	0.07 	   		& 1.14[0]   	& 6.24[7]  & 4.48[7]   & 11\%  		   \\
                    &                       &               &    417.16$^b$ &                      &               &          &           &              \\ 
2p4f	G(7/2)$_3$	&	2p3d	$^1$D$^o_2$	&	413.652 	&	413.294 	&	0.09      		& 4.08[-1]		& 2.27[7]  & 2.04[7]   & 13\%  	  	\\	
                    &                       &               &   413.18$^b$  &                       &               &          &           &             \\ 
2p4f	D(5/2)$_3$	&	2p3d	$^1$D$^o_2$	&	411.568 	&	411.199 	&	0.09 	   		& 5.34[-2]		&	3.00[6] &		    &				\\	
2p4f	F(5/2)$_3$	&	2p3d	$^3$D$^o_2$	&	423.694 	&	424.295 	&	$-0.14$ 	  	& 1.53[0]		&	8.12[7] &		    &			  	\\	
2p4f	F(7/2)$_3$	&	2p3d	$^3$D$^o_2$	&	423.258 	&	423.824 	&	$-0.13$ 	  	& 1.79[0]   	&	9.55[7] &		    &			  	\\	
2p4f	G(7/2)$_3$	&	2p3d	$^3$D$^o_2$	&	419.252 	&	419.715 	&	$-0.11$ 	    & 3.14[-1]		&	1.70[7] &		    &			    \\	
2p4f	D(5/2)$_3$	&	2p3d	$^3$D$^o_2$	&	417.111 	&	417.556 	&	$-0.11$ 	 	& 2.08[-3]		&	1.14[5] &		    &				\\	
2p4f	F(5/2)$_3$	&	2p3d	$^3$P$^o_2$	&	450.585 	&	451.005 	&	$-0.09$ 	    & 1.22[-1]		&	5.74[6] &		    &			    \\	
2p4f	F(7/2)$_3$	&	2p3d	$^3$P$^o_2$	&	450.093 	&	450.473 	&	$-0.08$ 	 	& 7.86[-4]		&	3.70[4] &		    &				\\
2p4f	D(5/2)$_3$	&	2p3d	$^3$P$^o_2$	&	443.147 	&	443.398 	&	$-0.06$         & 3.73[0]		&	1.81[8] &		    &			   \\	
2p4f	F(5/2)$_2$	&	2p3d	$^3$F$^o_3$	&	409.326 	&	408.798 	&	0.13 		 	& 2.63[-2]		&	2.10[6] &	     	&				\\	
2p4f	D(5/2)$_2$	&	2p3d	$^3$F$^o_3$	&	403.135 	&	402.509 	&	0.16 		 	& 1.30[-2]		&	1.07[6] &		    &				\\	
2p4f	D(3/2)$_2$	&	2p3d	$^3$F$^o_3$	&	401.937 	&	401.295 	&	0.16 		 	& 6.78[-3]		&	5.60[5] &		    &				\\	
2p4f	F(5/2)$_2$	&	2p3d	$^3$D$^o_3$	&	424.300 	&	424.790 	&	$-0.12$  	    & 6.15[-3]		&	4.56[5] &		    &				\\	
2p4f	D(5/2)$_2$	&	2p3d	$^3$D$^o_3$	&	417.652 	&	418.003 	&	$-0.08$  	    & 4.86[-2]		&	3.72[6] &		    &				\\	
2p4f	D(3/2)$_2$	&	2p3d	$^3$D$^o_3$	&	416.366 	&	416.694 	&	$-0.08$  	    & 4.19[-2]		&	3.23[6] &		    &				\\	
2p4f	F(5/2)$_2$	&	2p3d	$^1$F$^o_3$	&	461.115 	&	460.877 	&	0.05 	  	 	& 5.19[-4]		&	3.25[4] &		    &				\\
2p4f	D(5/2)$_2$	&	2p3d	$^1$F$^o_3$	&	453.274 	&	452.899 	&	0.08 	 	 	& 9.48[-3]		&	6.16[5] &		    &				\\	
2p4f	D(3/2)$_2$	&	2p3d	$^1$F$^o_3$	&	451.760 	&	451.363 	&	0.09 	 	 	& 9.85[-3]		&	6.44[5] &		    &				\\	
2p4f	F(5/2)$_3$	&	2p3d	$^3$F$^o_3$	&	409.351 	&	408.846 	&	0.12 	 	 	& 1.93[-1]		&	1.10[7] &		    &			   \\	
2p4f	F(7/2)$_3$	&	2p3d	$^3$F$^o_3$	&	408.944 	&	408.409 	&	0.13 	  	 	& 3.52[-2]		&	2.01[6] &		    &				\\	
2p4f	G(7/2)$_3$	&	2p3d	$^3$F$^o_3$	&	405.203 	&	404.592 	&   0.15         	& 4.50[-1]		&	2.61[7] & 2.14[7]	& 39\%			    \\
                    &                       &               &   404.48$^b$  &                       &               &          &           &                \\ 
2p4f	F(5/2)$_3$	&	2p3d	$^3$D$^o_3$	&	424.327 	&	424.842 	&	$-0.12$ 	   	& 7.84[-2]		&	4.15[6] &		    &			 	\\	
2p4f	F(7/2)$_3$	&	2p3d	$^3$D$^o_3$	&	423.891 	&	424.370 	&	$-0.11$ 		& 2.13[-1]		&	1.13[7] &		    &			   \\	
2p4f	G(7/2)$_3$	&	2p3d	$^3$D$^o_3$	&	419.872 	&	420.251 	&	$-0.09$ 	  	& 3.91[-2]		&	2.11[6] &		    &			 	\\	
2p4f	D(5/2)$_3$	&	2p3d	$^3$D$^o_3$	&	417.725 	&	418.085 	&	$-0.09$ 	    & 7.75[-1]  	&	4.23[7] & 4.70[7]   & 23\%  		    \\
                    &                       &               &   417.97$^b$  &                      &               &          &           &               \\ 
2p4f	F(5/2)$_3$	&	2p3d	$^1$F$^o_3$	&	461.147 	&	460.938 	&	0.05 	 	& 2.64[-1]		&	1.18[7] &		    &			    \\	
2p4f	F(7/2)$_3$	&	2p3d	$^1$F$^o_3$	&	460.631 	&	460.382 	&	0.05 		& 2.16[-1]		&	9.69[6] &		    &			    \\	
2p4f	G(7/2)$_3$	&	2p3d	$^1$F$^o_3$	&	455.890 	&	455.538 	&	0.08 	  	& 4.04[-2]		&	1.85[6] &		    &				\\	
2p4f	D(5/2)$_3$	&	2p3d	$^1$F$^o_3$	&	453.360 	&	452.995 	&	0.08 	 	& 1.73[-2]		&	8.03[5] &		    &				\\
2p4f	F(7/2)$_4$	&	2p3d	$^3$F$^o_3$	&	408.865 	&	408.342     &	0.13      	& 8.18[-1]		&	3.63[7] & 3.35[7]   & 16\%  		    \\
                    &                       &               &  408.23$^b$   &                      &               &          &           &               \\ 
2p4f	G(7/2)$_4$	&	2p3d	$^3$F$^o_3$	&	405.066 	&	404.467     &	0.15     	& 2.65[0]		&	1.20[8] & 1.25[8]   & 25\%  	     \\	
                    &                       &               &   404.35$^b$  &                      &               &          &           &               \\ 
2p4f	G(9/2)$_4$	&	2p3d	$^3$F$^o_3$	&	403.375 	&	402.722     &	0.16      	& 1.68[0]		&	7.65[7] & 6.72[7]   & 15\%  		     \\	
                    &                       &               &   402.61$^b$  &                      &               &          &           &                \\ 
2p4f	F(7/2)$_4$	&	2p3d	$^3$D$^o_3$	&	423.805 	&	424.298	    &	$-0.12$     	& 4.39[0]		&	1.81[8] &		    &				     \\	
                    &                       &               &   424.1$^c$   &                       &               &          &           &                \\ 
2p4f	G(7/2)$_4$	&	2p3d	$^3$D$^o_3$	&	419.725 	&	420.116     &	$-0.09$ 	    & 6.90[-1]		&	2.90[7] &		    &	    	     \\	
2p4f	G(9/2)$_4$	&	2p3d	$^3$D$^o_3$	&	417.910 	&	418.233     &	$-0.08$ 	    & 6.76[-3]		&	2.87[5] &		    &				\\	
2p4f	F(7/2)$_4$	&	2p3d	$^1$F$^o_3$	&	460.530 	&	460.297     &	0.05 	    & 2.27[-1]		&	7.95[6] &		    &				    \\	
2p4f	G(7/2)$_4$	&	2p3d	$^1$F$^o_3$	&	455.717 	&	455.380     &	0.07 	    & 1.91[0]   	&	6.80[7] & 6.11[7]   & 9\%   	  	\\
                    &                       &               &   455.25$^b$  &               &               &          &           &                \\ 
2p4f	G(9/2)$_4$	&	2p3d	$^1$F$^o_3$	&	453.577 	&	453.168     &	0.09 	  	& 4.20[0]		&	1.51[8] & 1.45[8]   & 20\%  		    \\	
                    &                       &               &   453.04$^b$  &               &               &          &           &               \\ 
2p4f	F(5/2)$_3$	&	2p3d	$^3$F$^o_4$	&	410.903 	&	410.213 	&	0.17 	  	& 8.97[-3]		&	5.06[5] &		    &				\\	
2p4f	F(7/2)$_3$	&	2p3d	$^3$F$^o_4$	&	410.493 	&	409.773 	&	0.18 	 	& 1.05[-2]		&	5.91[5] &		    &				\\	
2p4f	G(7/2)$_3$	&	2p3d	$^3$F$^o_4$	&	406.724 	&	405.931 	&	0.20 	 	& 1.26[-2]		&	7.23[5] &		    &				\\	
2p4f	D(5/2)$_3$	&	2p3d	$^3$F$^o_4$	&	404.708 	&	403.910 	&	0.20 	 	& 3.46[-2]		&	2.01[6] &		    &				\\
2p4f	F(7/2)$_4$	&	2p3d	$^3$F$^o_4$	&	410.413 	&	409.706 	&	0.17 	 	& 3.01[-1]		&	1.33[7] &		    &			    \\	
2p4f	G(7/2)$_4$	&	2p3d	$^3$F$^o_4$	&	406.586 	&	405.805 	&	0.19      	& 4.78[-1]  	&	2.14[7] & 1.99[7]   & 20\%  		     \\
                    &                       &               &   405.69$^b$  &               &               &          &           &                \\ %
2p4f	G(9/2)$_4$	&	2p3d	$^3$F$^o_4$	&	404.882 	&	404.048 	&	0.21 	 	& 1.08[-1]		&	4.90[6] &		    &			     \\	
2p4f	G(9/2)$_5$	&	2p3d	$^3$F$^o_4$	&	405.064 	&	404.245 	&	0.20 	    & 6.59[0]   	&	2.44[8] & 2.08[8]   & 10\%  	      \\	
                    &                       &               &   404.13$^b$  &               &               &          &           &               \\  \\
\multicolumn{9}{c}{2p4f -- 2p4s}\\
2p4f	D(3/2)$_1$	&	2p4s	$^3$P$^o_0$	&	668.378 	&  669.060    	&	$-0.10$ 	  	& 1.03[-2]		&	5.11[5] &	    	&				\\	
2p4f	D(3/2)$_1$	&	2p4s	$^3$P$^o_1$	&	670.747 	&  671.388  	&	$-0.10$ 	 	& 7.61[-3]		&	3.76[5] &	    	&				\\	
2p4f	D(5/2)$_2$	&	2p4s	$^3$P$^o_1$	&	673.855 	&	674.623     &	$-0.11$  	 	& 7.25[-3]		&	2.13[5] &		    &				\\	
2p4f	D(3/2)$_2$	&	2p4s	$^3$P$^o_1$	&	670.514 	&	671.221     &	$-0.11$  	 	& 1.57[-2]		&	4.65[5] &		    &				\\	
2p4f	F(5/2)$_2$	&	2p4s	$^1$P$^o_1$	&	758.779 	&	759.057     &	$-0.04$  	 	& 2.71[-3]		&	6.28[4] &		    &				\\	
2p4f	D(5/2)$_2$	&	2p4s	$^1$P$^o_1$	&	737.777 	&	737.655     &	0.02 	 	 	& 4.56[-2]		&	1.12[6] &		    &			    \\	
2p4f	D(3/2)$_2$	&	2p4s	$^1$P$^o_1$	&	733.774 	&	733.589     &	0.03 	 	 	& 3.73[-2]		&	9.25[5] &		    &				\\	
2p4f	D(3/2)$_1$	&	2p4s	$^3$P$^o_2$	&	676.249 	&	676.817     &	$-0.08$   	 	& 5.00[-4]		&	2.43[4] &		    &				\\
2p4f	D(5/2)$_2$	&	2p4s	$^3$P$^o_2$	&	679.408 	&	680.105 	&	$-0.10$ 	  	& 3.31[-3]		&	9.57[4] &		    &				\\	
2p4f	D(3/2)$_2$	&	2p4s	$^3$P$^o_2$	&	676.013 	&	676.647 	&	$-0.09$ 	 	& 4.01[-3]		&	1.17[5] &		    &				\\	
2p4f	F(5/2)$_3$	&	2p4s	$^3$P$^o_2$	&	697.252 	&	698.396 	&	$-0.16$ 	 	& 2.17[-3]		&	4.26[4] &		    &				\\	
2p4f	D(5/2)$_3$	&	2p4s	$^3$P$^o_2$	&	679.601 	&	680.322 	&	$-0.11$ 	 	& 3.96[-2]		&	8.16[5] &		    &				\\ \\
\multicolumn{9}{c}{2p4f -- 2p4d}\\
2p4f	D(3/2)$_1$	&	2p4d	$^3$P$^o_0$	&	12974.039 	& 14081.929 	&	$-7.87$ 	 	& 9.04[-2]		&	1.19[4] &		    &			     \\	
2p4f	D(3/2)$_1$	&	2p4d	$^3$D$^o_1$	&	7677.248 	& 8020.854	    &	$-4.28$ 	 	& 2.64[-2]		&	9.97[3] &		    &			    \\	
2p4f	D(3/2)$_1$	&	2p4d	$^3$P$^o_1$	&	12567.551 	& 13602.666 	&	$-7.61$ 	 	& 7.81[-2]		&	1.10[4] &		    &				    \\	
2p4f	F(5/2)$_2$	&	2p4d	$^3$D$^o_1$	&	11646.596 	&	12608.750   &	$-7.63$ 	 	& 2.88[-1]		&	2.83[4] &		    &				    \\	
2p4f	D(3/2)$_2$	&	2p4d	$^3$D$^o_1$	&	7646.897 	&	7996.993    &	$-4.38$ 	 	& 1.16[-2]		&	2.65[3] &		    &			    \\	
2p4f	D(5/2)$_2$	&	2p4d	$^3$P$^o_1$	&	13756.104 	&	15066.367   &	$-8.70$ 	 	& 9.77[-2]		&	6.89[3] &		    &			    \\	
2p4f	D(3/2)$_2$	&	2p4d	$^3$P$^o_1$	&	12486.265 	&	13534.181   &	$-7.74$ 		& 9.55[-2]		&	8.17[3] &		    &			    \\	
2p4f	D(5/2)$_2$	&	2p4d	$^1$P$^o_1$	&	132082.948 	&	126582.278  &	4.35 		 	& 1.16[-2]		&	8.86[0] &		    &			    \\	
2p4f	D(3/2)$_2$	&	2p4d	$^1$P$^o_1$	&	66827.052 	&	64876.087   &	3.01 	  		& 2.59[-2]		&	7.73[1] &		    &			    \\	
2p4f	D(3/2)$_1$	&	2p4d	$^3$F$^o_2$	&	5521.262 	&	5515.933    &	0.10 	  	 	& 1.14[-3]		&	8.32[2] &		    &				\\	
2p4f	D(3/2)$_1$	&	2p4d	$^3$D$^o_2$	&	7845.783 	&	8193.095    &	$-4.24$ 	 	& 1.06[-2]		&	3.82[3] &		    &			    \\	
2p4f	D(3/2)$_1$	&	2p4d	$^3$P$^o_2$	&	11891.879 	&	12798.853   &	$-7.09$ 	 	& 6.94[-3]		&	1.09[3] &		    &			    \\	
2p4f	F(5/2)$_2$	&	2p4d	$^3$F$^o_2$	&	7313.898 	&	7356.836 	&	$-0.58$ 	 	& 5.33[-2]		&	1.33[4] &		    &			    \\	
2p4f	D(5/2)$_2$	&	2p4d	$^3$F$^o_2$	&	5739.144 	&	5742.143 	&	$-0.05$ 	 	& 9.38[-4]		&	3.80[2] &		    &				\\
2p4f	D(3/2)$_2$	&	2p4d	$^3$F$^o_2$	&	5505.517 	&	5504.638 	&	0.02 	  	    & 5.14[-4]		&	2.26[2] &		    &				\\
2p4f	F(5/2)$_2$	&	2p4d	$^1$D$^o_2$	&	8843.531 	&	9032.037 	&	$-2.09$ 	  	& 8.16[-4]		&	1.39[2] &		    &				\\
2p4f	D(5/2)$_2$	&	2p4d	$^1$D$^o_2$	&	6640.371 	&	6714.113 	&	$-1.10$ 	  	& 5.03[-2]		&	1.52[4] &		    &				    \\	
2p4f	D(3/2)$_2$	&	2p4d	$^1$D$^o_2$	&	6329.635 	&	6391.655 	&	$-0.97$ 	 	& 4.94[-2]		&	1.65[4] &		    &				    \\	
2p4f	F(5/2)$_2$	&	2p4d	$^3$D$^o_2$	&	12038.910 	&	13039.680 	&	$-7.67$ 	 	& 4.14[-2]		&	3.81[3] &		    &			    \\	
2p4f	D(5/2)$_2$	&	2p4d	$^3$D$^o_2$	&	8293.181 	&	8702.311 	&	$-4.70$ 	  	& 4.50[-2]		&	8.73[3] &		    &				    \\	
2p4f	D(3/2)$_2$	&	2p4d	$^3$D$^o_2$	&	7814.087 	&	8168.200 	&	$-4.34$ 	 	& 1.28[-2]		&	2.79[3] &		    &				    \\	
2p4f	D(5/2)$_2$	&	2p4d	$^3$P$^o_2$	&	12950.852 	&	14086.491 	&	$-8.06$ 	  	& 3.46[-2]		&	2.75[3] &		    &				    \\	
2p4f	D(3/2)$_2$	&	2p4d	$^3$P$^o_2$	&	11819.213 	&	12738.204 	&	$-7.21$ 	 	& 5.30[-2]		&	5.06[3] &		    &				    \\	
2p4f	F(5/2)$_3$	&	2p4d	$^3$F$^o_2$	&	7321.984 	&	7372.348 	&	$-0.68$ 	 	& 5.86[-3]		&	1.04[3] &		    &			    \\	
2p4f	F(7/2)$_3$	&	2p4d	$^3$F$^o_2$	&	7194.089 	&	7232.698 	&	$-0.53$ 		& 2.80[-1]		&	5.15[4] &		    &			    \\	
2p4f	G(7/2)$_3$	&	2p4d	$^3$F$^o_2$	&	6188.808 	&	6197.400 	&	$-0.14$ 		& 6.17[-1]		&	1.53[5] &		    &			    \\	
2p4f	F(5/2)$_3$	&	2p4d	$^1$D$^o_2$	&	8855.278 	&	9055.428 	&	$-2.21$ 		& 3.83[-1]		&	4.65[4] &		    &			    \\	
2p4f	F(7/2)$_3$	&	2p4d	$^1$D$^o_2$	&	8668.892 	&	8845.644 	&	$-2.00$ 		& 1.24[-1]		&	1.57[4] &		    &			    \\	
2p4f	G(7/2)$_3$	&	2p4d	$^1$D$^o_2$	&	7249.844 	&	7345.002 	&	$-1.30$ 		& 1.45[-1]		&	2.62[4] &		    &			    \\	
2p4f	D(5/2)$_3$	&	2p4d	$^1$D$^o_2$	&	6658.854 	&	6735.322 	&	$-1.14$ 	  	& 4.35[-3]		&	9.36[2] &		    &				\\	
2p4f	F(5/2)$_3$	&	2p4d	$^3$D$^o_2$	&	12060.835 	&	13088.491 	&	$-7.85$ 	  	& 1.60[-1]		&	1.05[4] &		    &				    \\	
2p4f	F(7/2)$_3$	&	2p4d	$^3$D$^o_2$	&	11717.560 	&	12654.704 	&	$-7.41$ 	 	& 2.07[-1]		&	1.44[4] &		    &				    \\	
2p4f	G(7/2)$_3$	&	2p4d	$^3$D$^o_2$	&	9266.123 	&	9792.497 	&	$-5.38$ 	 	& 6.44[-2]		&	7.14[3] &		    &			    \\	
2p4f	D(5/2)$_3$	&	2p4d	$^3$D$^o_2$	&	8322.029 	&	8737.974 	&	$-4.76$ 		& 9.47[-3]		&	1.30[3] &		    &			    \\	
2p4f	F(5/2)$_3$	&	2p4d	$^3$P$^o_2$	&	25285.729 	&	30787.229 	&	$-17.87$ 	  	& 9.78[-4]		&	1.46[1] &		    &				\\
2p4f	D(5/2)$_3$	&	2p4d	$^3$P$^o_2$	&	13021.342 	&	14180.173 	&	$-8.17$ 	 	& 3.60[-1]		&	2.02[4] &		    &			    \\	
2p4f	F(5/2)$_2$	&	2p4d	$^3$F$^o_3$	&	7706.238 	&	7722.246 	&	$-0.21$ 	 	& 5.21[-3]		&	1.17[3] &		    &			   \\	
2p4f	D(5/2)$_2$	&	2p4d	$^3$F$^o_3$	&	5977.929 	&	5962.354 	&	0.26 	  	& 1.95[-3]		&	7.26[2] &		    &				\\	
2p4f	D(3/2)$_2$	&	2p4d	$^3$F$^o_3$	&	5724.918 	&	5706.688 	&	0.32 	 	& 8.42[-4]		&	3.43[2] &		    &				\\
2p4f	F(5/2)$_2$	&	2p4d	$^3$D$^o_3$	&	12610.818 	&	13675.214 	&	$-7.78$ 	  	& 8.28[-4]		&	6.95[1] &		    &				\\
2p4f	D(5/2)$_2$	&	2p4d	$^3$D$^o_3$	&	8560.691 	&	8980.853 	&	$-4.68$ 	  	& 8.14[-3]		&	1.48[3] &		    &				    \\	
2p4f	D(3/2)$_2$	&	2p4d	$^3$D$^o_3$	&	8051.076 	&	8413.118 	&	$-4.30$	 	& 7.22[-3]		&	1.49[3] &		    &				    \\	
2p4f	F(5/2)$_3$	&	2p4d	$^3$F$^o_3$	&	7715.216 	&	7739.339 	&	$-0.31$ 	 	& 3.75[-2]		&	6.00[3] &		    &				    \\	
2p4f	F(7/2)$_3$	&	2p4d	$^3$F$^o_3$	&	7573.291 	&	7585.584 	&	$-0.16$ 	 	& 4.98[-3]		&	8.27[2] &		    &				    \\	
2p4f	G(7/2)$_3$	&	2p4d	$^3$F$^o_3$	&	6467.427 	&	6454.694 	&	0.20 	  	& 1.00[-1]		&	2.28[4] &		    &				    \\	
2p4f	F(5/2)$_3$	&	2p4d	$^3$D$^o_3$	&	12634.877 	&	13728.909 	&	$-7.97$ 	 	& 9.12[-3]		&	5.45[2] &	        &			    \\	
2p4f	F(7/2)$_3$	&	2p4d	$^3$D$^o_3$	&	12258.808 	&	13252.405 	&	$-7.50$ 		& 2.58[-2]		&	1.63[3] &	        &			    \\	
2p4f	G(7/2)$_3$	&	2p4d	$^3$D$^o_3$	&	9601.260 	&	10146.619 	&	$-5.37$ 	  	& 8.20[-3]		&	8.48[2] &	        &				    \\	
2p4f	D(5/2)$_3$	&	2p4d	$^3$D$^o_3$	&	8591.361 	&	9018.840 	&	$-4.74$ 	 	& 1.22[-1]		&	1.58[4] &	        &			    \\	
2p4f	G(7/2)$_3$	&	2p4d	$^1$F$^o_3$	&	44006.337 	&	54466.231 	&	$-19.20$ 	 	& 1.28[-3]		&	6.28[0] &	        &			    \\	
2p4f	D(5/2)$_3$	&	2p4d	$^1$F$^o_3$	&	28599.211 	&	32590.275 	&	$-12.25$ 	  	& 1.05[-3]		&	1.22[1] &	        &	    		 	\\
2p4f	F(7/2)$_4$	&	2p4d	$^3$F$^o_3$	&	7546.031 	&	7562.752 	&	$-0.22$ 	  	& 2.02[-1]		&	2.63[4] &		    &				    \\	
2p4f	G(7/2)$_4$	&	2p4d	$^3$F$^o_3$	&	6432.730 	&	6423.103 	&	0.15 	  	& 6.09[-1]		&	1.09[5] &		    &				    \\	
2p4f	G(9/2)$_4$	&	2p4d	$^3$F$^o_3$	&	6031.145 	&	6009.399 	&	0.36 	 	& 3.84[-1]		&	7.82[4] &		    &				    \\	
2p4f	F(7/2)$_4$	&	2p4d	$^3$D$^o_3$	&	12187.542 	&	13182.873 	&	$-7.55$ 	 	& 4.66[-1]		&	2.33[4] &		    &			    \\	
2p4f	G(7/2)$_4$	&	2p4d	$^3$D$^o_3$	&	9524.989 	&	10068.770 	&	$-5.40$ 	 	& 1.13[-1]		&	9.22[3] &		    &			    \\	
2p4f	G(9/2)$_4$	&	2p4d	$^3$D$^o_3$	&	8670.245 	&	9088.017 	&	$-4.60$ 	  	& 6.58[-3]		&	6.49[2] &		    &				    \\	
2p4f	G(7/2)$_4$	&	2p4d	$^1$F$^o_3$	&	42450.227 	&	52295.785 	&	$-18.83$ 	 	& 5.01[-2]		&	2.06[2] &		    &			    \\	
2p4f	G(9/2)$_4$	&	2p4d	$^1$F$^o_3$	&	29491.565 	&	33512.064 	&	$-12.00$ 		& 1.73[-1]		&	1.48[3] &		    &			    \\
2p4f	F(5/2)$_3$	&	2p4d	$^3$F$^o_4$	&	8307.373 	&	8290.774 	&	0.20 	  	& 1.89[-3]		&	2.61[2] &		    &				\\	
2p4f	F(7/2)$_3$	&	2p4d	$^3$F$^o_4$	&	8143.124 	&	8114.578 	&	0.35 	 	& 2.00[-3]		&	2.87[2] &		    &				\\	
2p4f	G(7/2)$_3$	&	2p4d	$^3$F$^o_4$	&	6878.431 	&	6833.775 	&	0.65 	 	& 2.76[-3]		&	5.56[2] &		    &				\\	
2p4f	D(5/2)$_3$	&	2p4d	$^3$F$^o_4$	&	6344.171 	&	6302.945 	&	0.65 	 	& 5.94[-3]		&	1.41[3] &		    &			    \\
2p4f	F(7/2)$_4$	&	2p4d	$^3$F$^o_4$	&	8111.616 	&	8088.455 	&	0.29 	 	& 5.69[-2]		&	6.41[3] &		    &			    \\	
2p4f	G(7/2)$_4$	&	2p4d	$^3$F$^o_4$	&	6839.197 	&	6798.374 	&	0.60 	 	& 1.06[-1]		&	1.67[4] &		    &			    \\	
2p4f	G(9/2)$_4$	&	2p4d	$^3$F$^o_4$	&	6387.083 	&	6336.654 	&	0.80 		& 2.54[-2]		&	4.62[3] &	        &				 	\\
2p4f	G(9/2)$_5$	&	2p4d	$^3$F$^o_4$	&	6432.771 	&	6385.533 	&	0.74 		& 1.54[0]		&	2.25[5] &		    &				    \\	
\enddata
\end{deluxetable}


\clearpage

\clearpage


\begin{deluxetable}{rrrrrrr}
\tabletypesize{\scriptsize}
\tablecaption{\label{tab:table8}
{\footnotesize
$gf$ values scaled with experimental transition energies.
The number in the square bracket represents the power of 10.
}}
\tablewidth{0pt} \tabcolsep 5pt
\tablehead{
&& \multicolumn{2}{c}{  $\lambda$ (nm) }   &\multicolumn{3}{c}{ $gf$ } \\
\cline{3-4}\cline{6-7}
\colhead{Upper} &\colhead{Lower} &\colhead{Calc.} &\colhead{Obs.}&  &  \colhead{Calc.} & \colhead{Scale}
 }
\startdata
2p4f	G(7/2)$_3$	&	2p4d	$^1$F$^o_3$	&	44006.337	&	54466.231	&	&	1.28[-3]	&	1.03[-3]	\\
2p4f	G(7/2)$_4$	&	2p4d	$^1$F$^o_3$	&	42450.227	&	52295.785	&	&	5.01[-2]	&	4.07[-2]	\\
2p4f	F(5/2)$_3$	&	2p4d	$^3$P$^o_2$	&	25285.729	&	30787.229	&	&	9.78[-4]	&	8.04[-4]	\\
2p4f	D(5/2)$_3$	&	2p4d	$^1$F$^o_3$	&	28599.211	&	32590.275	&	&	1.05[-3]	&	9.20[-4]	\\
2p4f	G(9/2)$_4$	&	2p4d	$^1$F$^o_3$	&	29491.565	&	33512.064	&	&	1.73[-1]	&	1.53[-1]	\\
2p4f	D(5/2)$_2$	&	2p4d	$^3$P$^o_1$	&	13756.104	&	15066.367	&	&	9.77[-2]	&	8.92[-2]	\\
2p4f	D(5/2)$_3$	&	2p4d	$^3$P$^o_2$	&	13021.342	&	14180.173	&	&	3.60[-1]	&	3.31[-1]	\\
2p4f	D(5/2)$_2$	&	2p4d	$^3$P$^o_2$	&	12950.852	&	14086.491	&	&	3.46[-2]	&	3.18[-2]	\\
2p4f	F(5/2)$_3$	&	2p4d	$^3$D$^o_3$	&	12634.877	&	13728.909	&	&	9.12[-3]	&	8.40[-3]	\\
2p4f	D(3/2)$_1$	&	2p4d	$^3$P$^o_0$	&	12974.039	&	14081.929	&	&	9.04[-2]	&	8.33[-2]	\\
2p4f	F(5/2)$_3$	&	2p4d	$^3$D$^o_2$	&	12060.835	&	13088.491	&	&	1.60[-1]	&	1.47[-1]	\\
2p4f	F(5/2)$_2$	&	2p4d	$^3$D$^o_3$	&	12610.818	&	13675.214	&	&	8.29[-4]	&	7.64[-4]	\\
2p4f	D(3/2)$_2$	&	2p4d	$^3$P$^o_1$	&	12486.265	&	13534.181	&	&	9.55[-2]	&	8.81[-2]	\\
2p4f	F(5/2)$_2$	&	2p4d	$^3$D$^o_2$	&	12038.91	&	13039.68	&	&	4.14[-2]	&	3.82[-2]	\\
2p4f	F(5/2)$_2$	&	2p4d	$^3$D$^o_1$	&	11646.596	&	12608.75	&	&	2.88[-1]	&	2.66[-1]	\\
2p4f	D(3/2)$_1$	&	2p4d	$^3$P$^o_1$	&	12567.551	&	13602.666	&	&	7.81[-2]	&	7.22[-2]	\\
2p4f	F(7/2)$_4$	&	2p4d	$^3$D$^o_3$	&	12187.542	&	13182.873	&	&	4.66[-1]	&	4.31[-1]	\\
2p4f	F(7/2)$_3$	&	2p4d	$^3$D$^o_3$	&	12258.808	&	13252.405	&	&	2.58[-2]	&	2.38[-2]	\\
2p4f	F(7/2)$_3$	&	2p4d	$^3$D$^o_2$	&	11717.56	&	12654.704	&	&	2.07[-1]	&	1.92[-1]	\\
2p4f	D(3/2)$_2$	&	2p4d	$^3$P$^o_2$	&	11819.213	&	12738.204	&	&	5.30[-2]	&	4.92[-2]	\\
2p4f	D(3/2)$_1$	&	2p4d	$^3$P$^o_2$	&	11891.879	&	12798.853	&	&	6.94[-3]	&	6.45[-3]	\\
2p4f	G(7/2)$_4$	&	2p4d	$^3$D$^o_3$	&	9524.989	&	10068.77	&	&	1.13[-1]	&	1.07[-1]	\\
2p4f	G(7/2)$_3$	&	2p4d	$^3$D$^o_2$	&	9266.123	&	9792.497	&	&	6.44[-2]	&	6.09[-2]	\\
2p4f	G(7/2)$_3$	&	2p4d	$^3$D$^o_3$	&	9601.26	    &	10146.619	&	&	8.20[-3]	&	7.76[-3]	\\
2p4f	D(5/2)$_3$	&	2p4d	$^3$D$^o_2$	&	8322.029	&	8737.974	&	&	9.47[-3]	&	9.02[-3]	\\
2p4f	D(5/2)$_3$	&	2p4d	$^3$D$^o_3$	&	8591.361	&	9018.84	    &	&	1.23[-1]	&	1.17[-1]	\\
2p4f	D(5/2)$_2$	&	2p4d	$^3$D$^o_2$	&	8293.181	&	8702.311	&	&	4.50[-2]	&	4.29[-2]	\\
2p4f	D(5/2)$_2$	&	2p4d	$^3$D$^o_3$	&	8560.691	&	8980.853	&	&	8.14[-3]	&	7.76[-3]	\\
2p4f	G(9/2)$_4$	&	2p4d	$^3$D$^o_3$	&	8670.245	&	9088.017	&	&	6.58[-3]	&	6.28[-3]	\\
2p4f	D(3/2)$_2$	&	2p4d	$^3$D$^o_1$	&	7646.897	&	7996.993	&	&	1.16[-2]	&	1.11[-2]	\\
2p4f	D(5/2)$_2$	&	2p4d	$^1$P$^o_1$	&	132082.948	&	126582.278	&	&	1.16[-2]	&	1.21[-2]	\\
2p4f	D(3/2)$_2$	&	2p4d	$^3$D$^o_2$	&	7814.087	&	8168.2	    &	&	1.28[-2]	&	1.22[-2]	\\
2p4f	D(3/2)$_2$	&	2p4d	$^3$D$^o_3$	&	8051.076	&	8413.118	&	&	7.22[-3]	&	6.91[-3]	\\
2p4f	D(3/2)$_1$	&	2p4d	$^3$D$^o_1$	&	7677.248	&	8020.854	&	&	2.64[-2]	&	2.53[-2]	\\
2p4f	D(3/2)$_1$	&	2p4d	$^3$D$^o_2$	&	7845.783	&	8193.095	&	&	1.06[-2]	&	1.01[-2]	\\
2p4f	D(3/2)$_2$	&	2p4d	$^1$P$^o_1$	&	66827.052	&	64876.087	&	&	2.59[-2]	&	2.67[-2]	\\
\enddata
\end{deluxetable}


\clearpage

\section{CONCLUSIONS}

We calculated the wavelengths and oscillator strengths for the transitions from the 2p4f configuration in N$^{+}$ using the GRASP2K package based on the multiconfiguration Dirac-Hartree-Fock method.
In order to deal with the pair-coupling level structure higher-order electron correlation effects were taken into account through an extended set of reference configurations. 
Also, the Breit interaction was included to improve fine structure splittings of the 2p4f configuration.
Except for some transitions with large wavelengths, uncertainties of present calculations were controlled within $3\%$ and $5\%$ for wavelengths and oscillator strengths, respectively.
We also compared our results with other theoretical and experimental values when available.
It was shown that previous calculations within the non-relativistic framework are not well suited for the level structure of the 2p4f configuration. Therefore, we recommended present results based on a fully relativistic method for abundance analysis and plasma diagnosis.
%

\acknowledgments
X. Z. Shen thanks Prof. P. Yuan for the discussion and the financial support by the Research and Development Program for Science and Technology of Hebei Province (Grant No.11217168), the Research and Development Program for Science and Technology of Handan (Grant No.1128103071, 1121120069-5), the Foundation of Handan College
(Grant No.09005).
J. G. Li and J. G. Wang were supported by the National Basic Research program of China under Grant No. 2013CB922200, and the National Science Foundation of China under Grant No. 11025417. 
%
%
%



\end{document}